\documentclass[%
 aip,
 amsmath,amssymb,
 reprint,%
groupedaddress
]{revtex4-1}

\usepackage{graphicx}% Include figure files
\usepackage{dcolumn}% Align table columns on decimal point
\usepackage{bm}% bold math
\usepackage[utf8]{inputenc}
\usepackage[T1]{fontenc}
\usepackage{mathptmx}
\usepackage{etoolbox}
\usepackage{amsthm}
\usepackage{xcolor}
\usepackage{subcaption}
\makeatletter
\def\@email#1#2{%
 \endgroup
 \patchcmd{\titleblock@produce}
  {\frontmatter@RRAPformat}
  {\frontmatter@RRAPformat{\produce@RRAP{*#1\href{mailto:#2}{#2}}}\frontmatter@RRAPformat}
  {}{}
}%
\makeatother

\newtheorem*{theorem*}{Framework}

\begin{document}

\preprint{AIP/123-QED}
%TC:ignore
\title{Symmetry of the critical current function in superconducting nanodevices}
% Force line breaks with \\
\author{Ziqi Zhao}
\author{Cliff Sun}
\author{Ci-You Huang}
\author{Jiankun Zhang}
\author{Xiangyu Song}
\author{Alexey Bezryadin}%
\affiliation{University of Illinois at Urbana-Champaign, Department of Materials Science and Engineering}
\affiliation{University of Illinois at Urbana-Champaign, Department of Physics}

\begin{abstract}
We study a variety of nano-scale superconducting devices containing more than one weak link. Even in devices with multiple weak links the IB symmetry is usually obeyed, which means that if we reverse both the bias current direction and the magnetic field direction at the same time, the superconducting response,  namely the critical current, remains unchanged. We also provide a detailed analysis of the situations in which such symmetry is violated.
\end{abstract}
%TC:endignore
\maketitle

%\tableofcontents

\section{Introduction}
Superconducting weak-link devices provide a sensitive platform for studying quantum phenomena in gate-tunable and hybrid qubit designs~\cite{koch_transmon_2007,larsen_nanowire_qubit_2015,de_lange_nanowire_josephson_qubit_2015,casparis_gatemon_2018}, fluxoid- and vorticity-based superconducting memory elements~\cite{murphy_nanowire_memory_2017,ilin_kinetic_inductance_memory_2021,golod_vortex_memory_2023}, and nonreciprocal superconducting transport~\cite{ando_sde_2020,baumgartner_supercurrent_rectification_2022,bauriedl_nbse2_diode_2022,song-2023}.
In conventional superconducting quantum interference devices (SQUIDs), the critical current $I_c$ oscillates with magnetic field because the superconducting phase differences across parallel weak links is constrained by the fluxoid quantization and the requirement that the wave function of the condensate is single-valued. 
This interference effect is widely used for magnetic-field sensing and also provides a direct probe of the current--phase relation, phase winding, and vortex configurations in mesoscopic superconducting circuits~\cite{tinkham_book, LikharevWeakLinks, little_parks, AndersonDayem}.

Nanowire-based SQUIDs differ from conventional tunnel-junction SQUIDs in an important way. 
A tunnel Josephson junction is characterized by a sinusoidal current--phase relation, whereas a thin superconducting nanowire or Dayem-bridge-like weak link can exhibit an approximately linear current--phase relation over a broad phase range, possibly much larger than $2\pi$. 
As a result, superconducting nanowire SQUIDs can display highly non-sinusoidal and multi-valued $I_c(B)$ patterns, reflecting the existence of multiple vorticity states rather than a single smooth interference envelope~\cite{model2}. 

Recent modeling of multiple-nanowire SQUIDs has shown\cite{model1} that the critical-current envelope can remain symmetric under simultaneous inversion of current and magnetic field, even when the detailed modulation pattern is complex and the devices involves many weak links. 
This symmetry, denoted here as $IB$ symmetry, can be written as
\begin{equation}
I_{c,+}(B) = -I_{c,-}(-B),
\end{equation}
where $I_{c,+}(B)$ and $I_{c,-}(B)$ are the positive and negative switching-current branches, respectively.

Understanding when this $IB$ symmetry is preserved or broken is important for two reasons. 
First, it provides a stringent diagnostic of whether the observed $I_c(B)$ envelope is governed by coherent phase constraints and vorticity-state selection, rather than by trivial field offsets, measurement artifacts, or uncontrolled trapped flux. 
Second, broken $IB$ symmetry is closely related to superconducting nonreciprocity, including the superconducting diode effect, where the critical current depends on the direction of current flow~\cite{AndoNature, BauriedlNatComm, song-2023}. 
Such nonreciprocal superconducting responses are of interest for dissipationless rectification, superconducting logic, magnetic-field sensing, and future superconducting or quantum-circuit architectures.

In this work, we experimentally study the $IB$ symmetry in several superconducting weak-link devices. 
The devices A and B are multi-nanowire weak-link structures fabricated using suspended SiN bridges acting as templates for various superconducting metals (Al or Ta). The device C is an Ag-nanowire templates coated with a superconducting Al film, thus making a hybrid bilayer nanowire\cite{bilayer}. 
These devices (A--C) exhibit different $I_c(B)$ modulation patterns, including triangular, irregular, and broad-lobe structures, yet all preserve the symmetry relation $I_{c,+}(B) = -I_{c,-}(-B)$ within experimental scatter. 
In contrast, hybrid tunnel-junction--nanobridge devices show clear violations of this symmetry. 
For Device D, a hybrid device consisting of a Sn tunnel junction and a nanobridge, we observe two types of symmetry breaking. In the first type, the two switching-current branches occur at nearly the same magnetic-field positions but have different current values. In the second type, the branches are shifted along the magnetic-field axis and also show different envelope shapes.
Device E, which is a proximized square array of Nb islands on a topological insulator film (BiSrTe), provides an additional example in which the symmetry breaking is observed. In this case it happens to concentrate near the central critical-current peak~\cite{song-2023}.

To interpret these observations, we develop and analyze models of tunnel-junction arrays, multiple-nanowire SQUIDs, and hybrid devices involving parallel connections of various Josephson junctions (JJ), namely nanobridges and superconductor-insulator-superconductor (SIS) junctions. 
We find that the models involving multiple-nanowire parallel weak links preserve $IB$ symmetry, but only when the full set of vorticity states is considered. the same is true for hybrid SIS--metallic-link devices. On the other hand, the hybrid SIS--metallic-link structure can introduce a bias-current-polarity-dependent energetic preference among competing vorticity states. 

This provides a possible microscopic mechanism for the observed symmetry breaking: the positive and negative switching-current branches can correspond to different metastable or ground-state vorticity configuration in the multivalued energy landscape.

\section{Method}
\subsection{Device Fabrication}
All devices were fabricated using either suspended nanobridges on $SiN$ membranes (Devices A–C) or shadow-mask-defined Sn junction structures on sapphire (Devices D–E). Device A was realized on a 
$\sim$ 200 nm-wide suspended $SiN$ nanobridge, formed by LPCVD, using $SiN$ and $SiO_2$ bilayer films on Si substrates\cite{Bezryadin1997}, e-beam lithography patterning, $SF_6$ dry etching, and HF undercut formation of the underlying $SiO_2$, followed by thermal evaporation deposition of 15 nm Al films to form superconducting weak links. The Device B used the same SiN-bridge platform, but was coated with 20nm Ta since Ta is a material of interest for qubit fabrication\cite{Place2021}. 

\begin{figure}[htbp]
  \centering

  \begin{subfigure}[t]{0.49\linewidth}
    \centering
    \includegraphics[width=\linewidth]{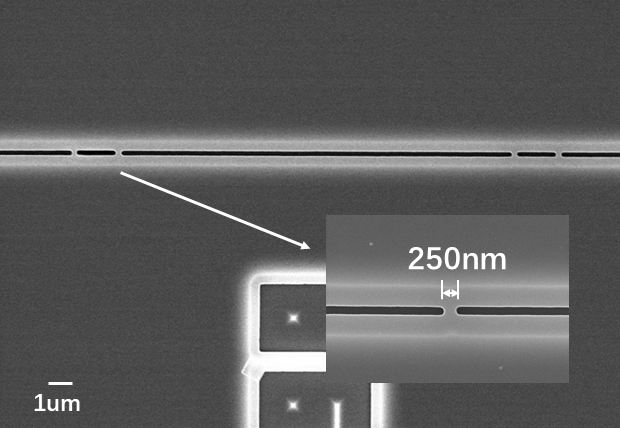}
    \caption{Device A}
    \label{fig:device_a}
  \end{subfigure}\hfill
  \begin{subfigure}[t]{0.49\linewidth}
    \centering
    \includegraphics[width=\linewidth]{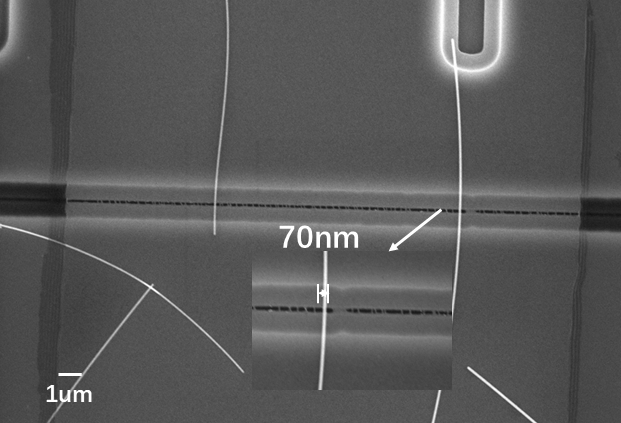}
    \caption{Device C}
    \label{fig:device_c}
  \end{subfigure}

  \caption{SEM images of the superconducting nanowire sample. a) A scanning electron microscope (SEM) image of the silicon nitride nanobridges in Device A. b) An SEM image of the silver nanowires in Device C.}
  \label{fig:sem}
\end{figure}

The Device C was fabricated by positioning a 70 nm-diameter Ag nanowire (from Novarials) across pre-etched trenches in SiN films. Thus, the Ag nanowire acts as a suspended template and it is coated by thermal evaporation of 10 nm Al. Thus produced Ag/Al bilayer nanowire weak links are superconducting. All devices were inspected by SEM to confirm successful suspension and film continuity prior to low-temperature transport measurements. Fig.~\ref{fig:sem} shows the SEM images of Devices A and C.

\begin{figure}[htbp]
  \centering
  \includegraphics[width=1\linewidth]{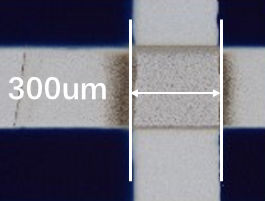}
  \caption{Optical microscope image of Device D, which shows the SIS junction made of Sn.}
  \label{fig:device_d}
\end{figure}

The Device D (on sapphire) was a hybrid Sn cross-junction: 100 nm Sn was first thermally evaporated through a 300 µm shadow mask, followed by $\sim$ 4 h ambient oxidation to form a native tin oxide tunnel barrier. The second Sn layer (200 nm) was deposited through a 300 µm slit mask to complete the cross geometry. This is shown in the optical microscope image in Fig.~\ref{fig:device_d}.

The Device E is a Nb--BST--Nb superconducting island array fabricated on a 40-nm-thick topological-insulator thin film with nominal composition $\mathrm{Bi}_{0.8}\mathrm{Sb}_{1.2}\mathrm{Te}_3$ (BST). This is the same sample as the one reported by Song \textit{et al.}~\cite{song-2023} in the superconducting diode effect investigation. The BST film was grown by molecular beam epitaxy and was chosen because this composition provides an undoped topological-insulator channel dominated by topological surface-state carriers. The device consists of a $23 \times 23$ square lattice of Nb islands deposited on the BST film by electron-beam lithography and plasma sputtering. Each Nb island has a thickness of 30 nm, an approximate lateral width of $w \approx 1.15~\mu\mathrm{m}$, and a nearest-neighbor gap of $d \approx 150~\mathrm{nm}$, giving a lattice constant of $w+d = 1.3~\mu\mathrm{m}$. This device geometry and material system are taken from Song.~\cite{song-2023}.
The device parameters are summarized in Table~\ref{tab:devices}.

\begingroup
\squeezetable
\begin{table}[htbp]
\begin{ruledtabular}
\begin{tabular}{ccccc}
Device name & Metal & Length  & Width & $I_c$ \\ 
\hline
A & Al & 200nm & 250nm & 385uA \\
B & Ta & 200nm & 250nm & 24uA \\
C & Al/Ag & 200nm & 70nm & 37uA \\
D & Sn & 300um & 300um & 700uA \\
E & Nb/BST & 1150nm & 1150nm & 2.8uA \\
\end{tabular}
\end{ruledtabular}

\caption{Summary of the device materials, nominal dimensions, and zero-field switching currents 
for the superconducting weak-link devices studied in this work. 
Devices A and B are superconducting nanowires templated by Silicon Nitride nanobridges, Device C is an 
Al-coated Ag nanowire, Device D is a Sn-based Josephson-junction, and Device E is a 
Nb/BST array of SNS (superconductor-normal metal- superconductor) junctions.}
\label{tab:devices}
\end{table}
\endgroup

\subsection{Experimental Setup}
Low-temperature transport measurements were performed in a cryogen-free He-3 refrigerator at a base temperature of $T_{base}\sim350~\mathrm{mK}$ using a four-probe configuration optimized for superconducting devices. A National Instruments NI-DAQ USB-6216 provided the bias by outputting an AC voltage $U$ to the device in series with a calibrated standard resistor $R_{\mathrm{st}}\approx 5~\mathrm{k}\Omega$. The NI-DAQ simultaneously recorded the voltage across the sample, $V$, and across the standard resistor, $V_{\mathrm{st}}$, at $\sim 10^{5}$ samples/s, with both signals preamplified (with PAR113) prior to digitization. The bias current was computed in real time as $I = V_{\mathrm{st}}/R_{\mathrm{st}}$. The V versus I dependence is displayed in LabVIEW and can be simultaneously analyzed to find zero bias resistance and the critical current. The amplitude of $U$ was chosen such that $I_{\max}>I_c$ for full switching characterization, while small-$U$ excitation was used for linear-response resistance measurements. For $I_c(B)$ measurements, a tunable magnetic field was generated by a suspended copper-wire solenoid with a calibrated conversion factor of $5.2~\mathrm{G/V}$, driven by a Keithley 2200 current source. Four filtered BNC lines (1.9~MHz low-pass) were used to suppress high-frequency noise. All measurement lines are filter with Cu powder filters installed at the base temperature. The solenoid voltage was recorded by the NI-DAQ and converted to magnetic field (Gauss), enabling continuous field sweeps and real-time acquisition of $I_c$ versus magnetic field in a stable, low-noise cryogenic environment. Previously, a sensitivity of this setup at the level of about 10 nT has been demonstrated~\cite{song-2023}.

\section{Experimental Results}

\subsection{Preservation of symmetry}

To characterize the field dependence of the superconducting critical current, we detect switching current events at various magnetic fields,  using a current-bias ramp protocol at $T=T_{base}$. For each magnetic field value $B$, the bias current is swept through the superconducting state until a switching event is detected when the device voltage exceeds a preset threshold $V_{\mathrm{th}}$. We define $I_{c,+}(B)$ as the positive switching current and $I_{c,-}(B)$ as the negative switching current (reported as negative values). In general, the raw curves $I_{c,+}(B)$ and $I_{c,-}(B)$ do not coincide because they correspond to opposite current polarities; therefore, to test the $I$--$B$ inversion symmetry of the critical-current envelopes (hereafter IB symmetry), we compare the positive branch $I_{c,+}(B)$ to the transformed negative branch $-I_{c,-}(-B)$. 

\begin{figure}[htbp]
  \centering

  \begin{subfigure}[t]{1\linewidth}
    \centering
    \includegraphics[width=\linewidth]{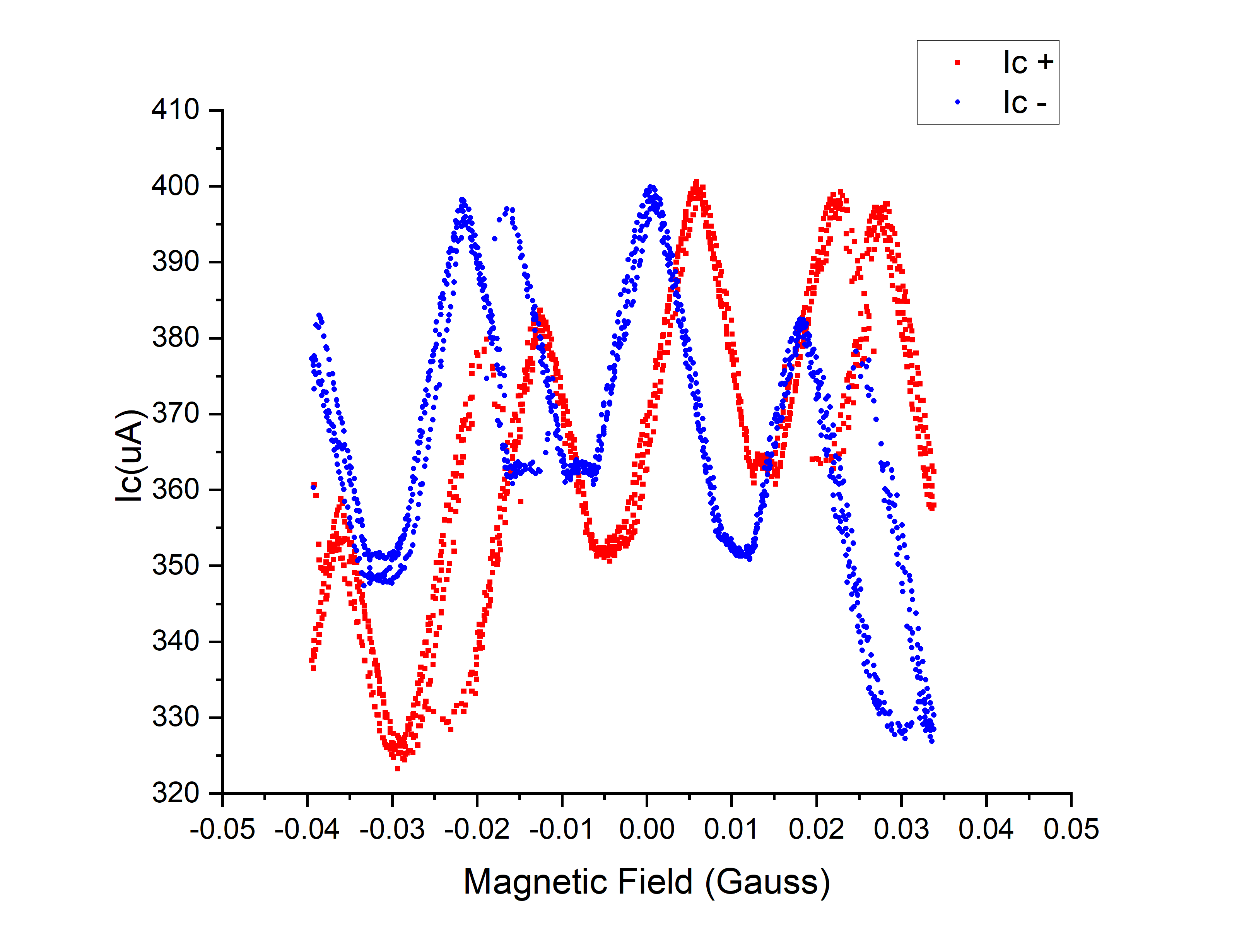}
    \caption{}
    \label{fig:device_a_IB_raw}
  \end{subfigure}\hfill
  \begin{subfigure}[t]{1\linewidth}
    \centering
    \includegraphics[width=\linewidth]{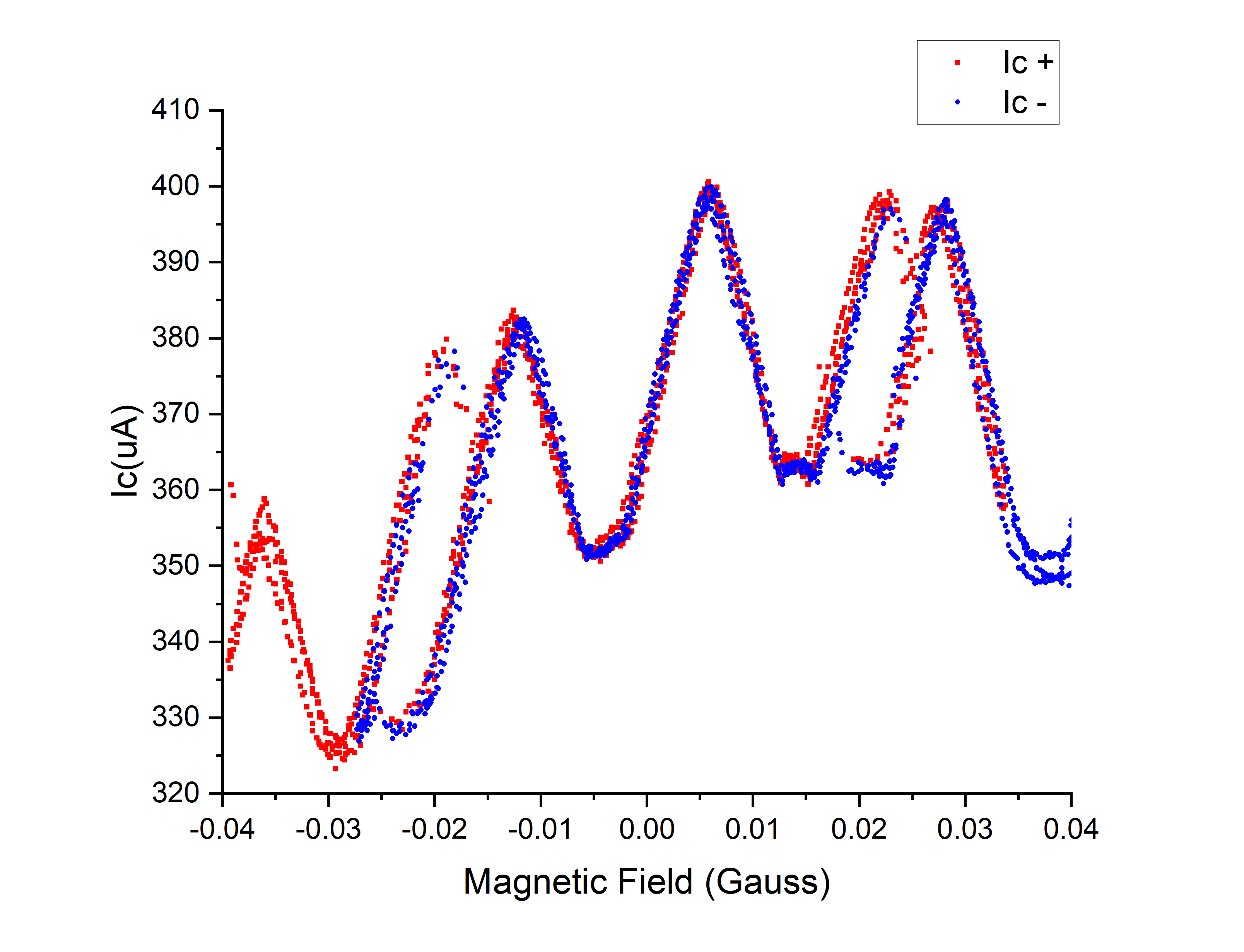}
    \caption{}
    \label{fig:device_a_IB_symmetry}
  \end{subfigure}

  \caption{Magnetic-field dependence of the switching-current envelopes for Device A. 
  (a) Comparison between the positive switching current $I_{c,+}(B)$ and the magnitude of the negative switching current, $-I_{c,-}(B)$. 
  (b) Comparison between $I_{c,+}(B)$ and the field-reversed negative branch, $-I_{c,-}(-B)$, used to examine $IB$ symmetry.}
  \label{fig:device_a_IB}
\end{figure}

Fig.~\ref{fig:device_a_IB}a presents a critical current versus magnetic field measurement taken on Device A. representative symmetry-preserved dataset for Device A. The red curve is the positive critical current and the blue curve is the absolute value of the negative critical current. Both of them are multi-valued. This reflects the fluctuations of the vorticity distribution in each new cooling of the nanowires, i.e., in each new measurement of the V-I curve. Within the measured field range $|B| < 0.04~\mathrm{G}$, both $I_{c,+}(B)$ and $I_{c,-}(B)$ show a pronounced nonmonotonic modulation with approximately linear segments as well as inverse-V-shape peak.
The modulation contains multiple local maxima and minima, with a characteristic spacing of approximately $0.02~\mathrm{G}$ between adjacent repeating features. 

We test the IB-symmetry in Fig.~\ref{fig:device_a_IB}a, where the magnetic field is inverted for the blue curve but it is kept unmodified for the red curve. This multi-bridge sample provides a particularly sensitive test for the IB-symmetry verification because each curve is not symmetric and not periodic, and because each curve shows a complex pattern for different height maxima and minima and because each curve is a multi-valued function.

Accordingly, we compare $I_{c,+}(B)$ with the transformed negative branch $-I_{c,-}(-B)$, as shown in Fig.~\ref{fig:device_a_IB}b. 
The two traces overlap closely over the full field range, including the sharp extrema and the multi-valued segments demonstrating that
\begin{equation}
I_{c,+}(B) = -I_{c,-}(-B)
\label{eq:ib_symmetry}
\end{equation}
Such result is in agreement with the symmetry predictions of Ref.\cite{model1}. This result also demonstrates that IB-symmetry is valid even if the critical current versus magnetic field function is multi-valued.

\begin{figure}[htbp]
  \centering

  \begin{subfigure}[t]{1\linewidth}
    \centering
    \includegraphics[width=0.9\linewidth]{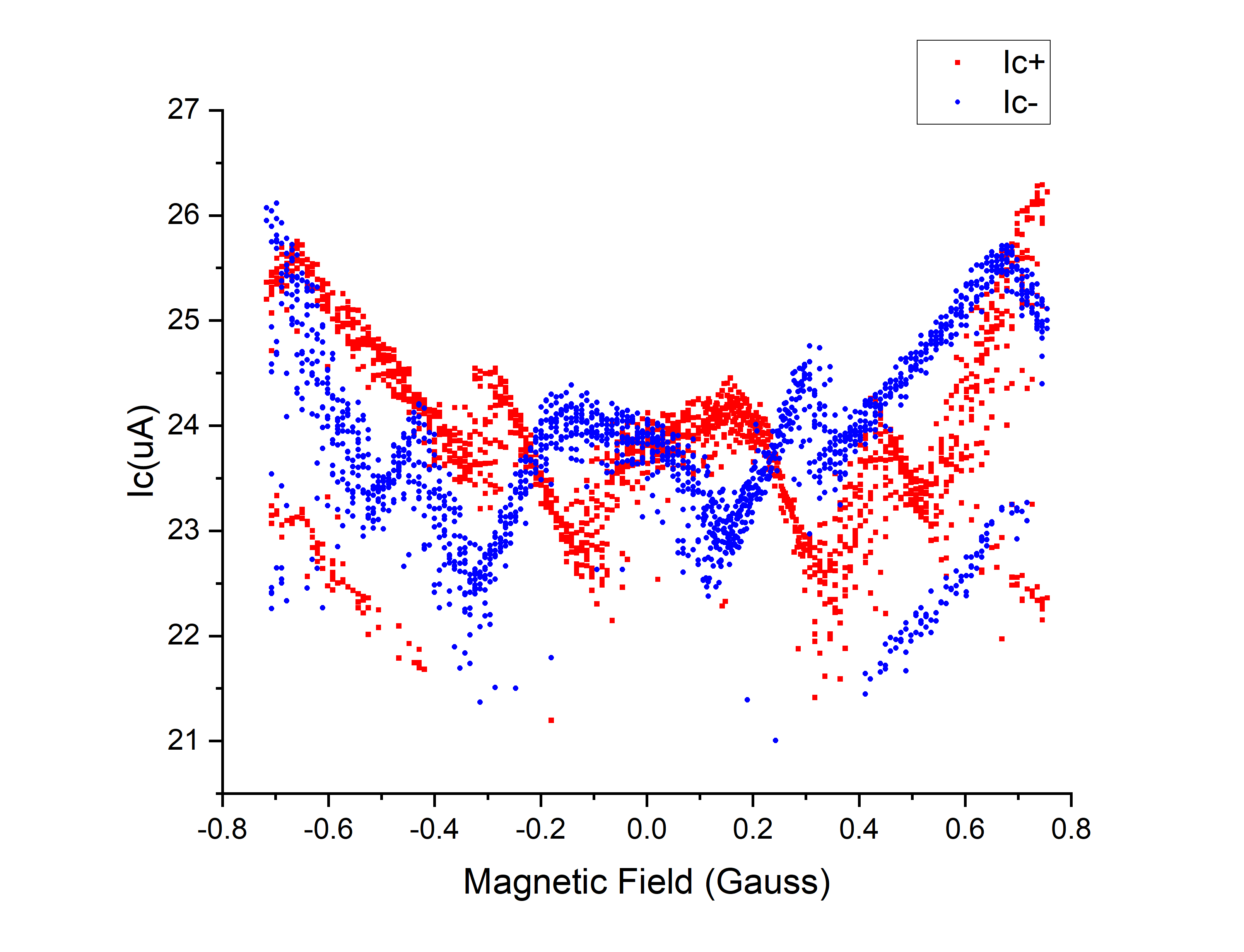}
    \caption{}
    \label{fig:device_b_IB_raw}
  \end{subfigure}

  \vspace{0.3cm}

  \begin{subfigure}[t]{1\linewidth}
    \centering
    \includegraphics[width=0.9\linewidth]{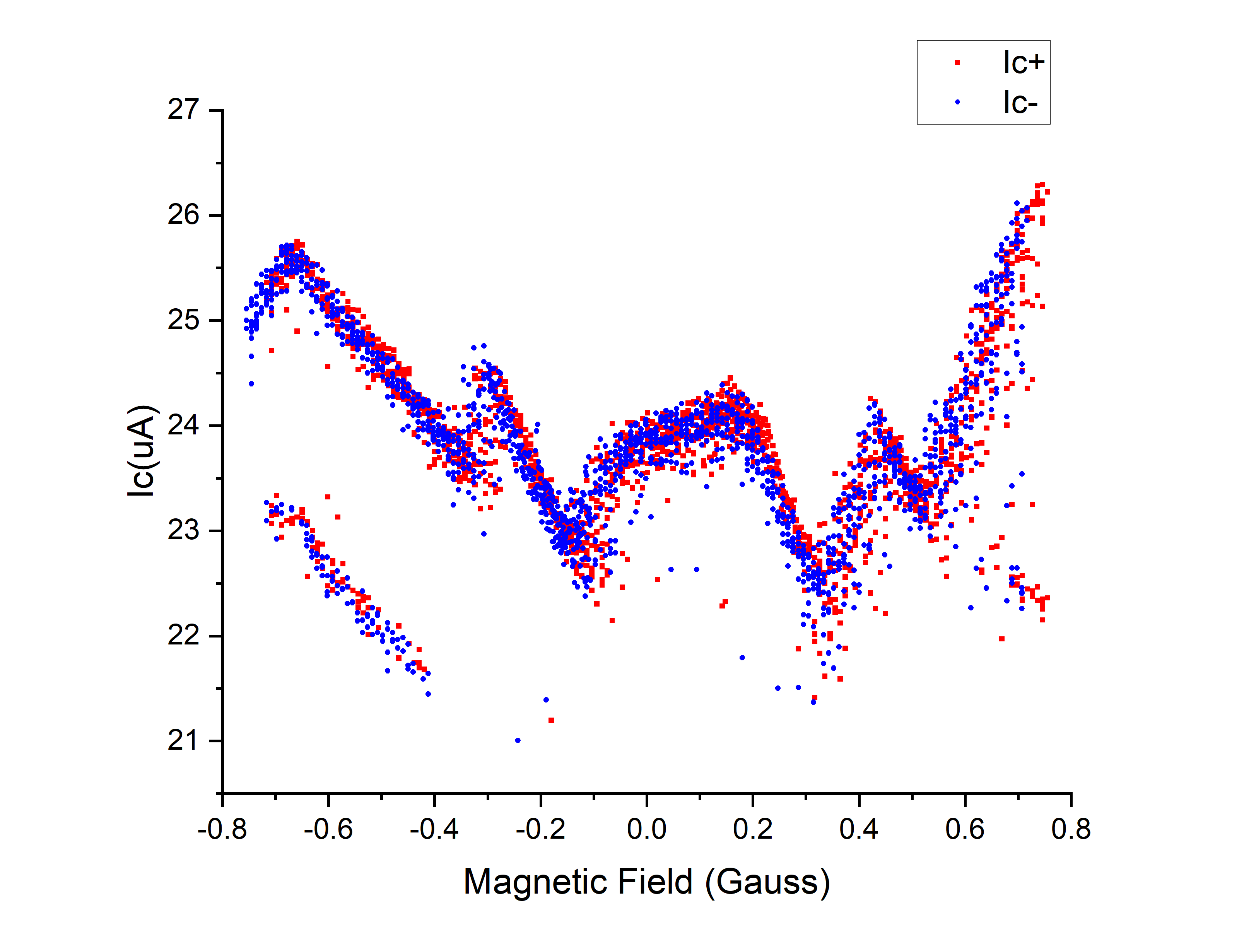}
    \caption{}
    \label{fig:device_b_IB_symmetry}
  \end{subfigure}

  \caption{Magnetic-field dependence of the switching-current (critical currents) for Device B. The curve is obtained by multiple repetitive measurements of the critical current of the device.
  Each dot represents one measured switching-current value at a given magnetic field.
  (a) Comparison between the positive switching current $I_{c,+}(B)$ and the magnitude of the negative switching current, $-I_{c,-}(B)$. 
  (b) Comparison between $I_{c,+}(B)$ and the field-reversed negative branch, $-I_{c,-}(-B)$, showing that Device B approximately preserves $IB$ symmetry.}
  \label{fig:device_b_IB}
\end{figure}

A similar symmetry-preserved behavior is observed in Device B shown in Fig.~\ref{fig:device_b_IB}. 
Unlike Device A, whose switching-current envelope shows an approximately triangular modulation, Device B exhibits a more irregular, nonmonotonic pattern over the measured range $|B| < 0.4~\mathrm{G}$. 
The envelope contains alternating upward and downward features, with several local peaks and dips of varying amplitude rather than a single repeating triangular form. 
Nevertheless, $I_{c,+}(B)$ closely matches the transformed branch $-I_{c,-}(-B)$, showing that the detailed modulation pattern is preserved under simultaneous inversion of current and magnetic field. 
This agreement demonstrates that Device B also satisfies the $IB$ symmetry relation defined in Eq.~\ref{eq:ib_symmetry}, within the experimental switching-current scatter.

\begin{figure}[htbp]
  \centering

  \begin{subfigure}[t]{1\linewidth}
    \centering
    \includegraphics[width=\linewidth]{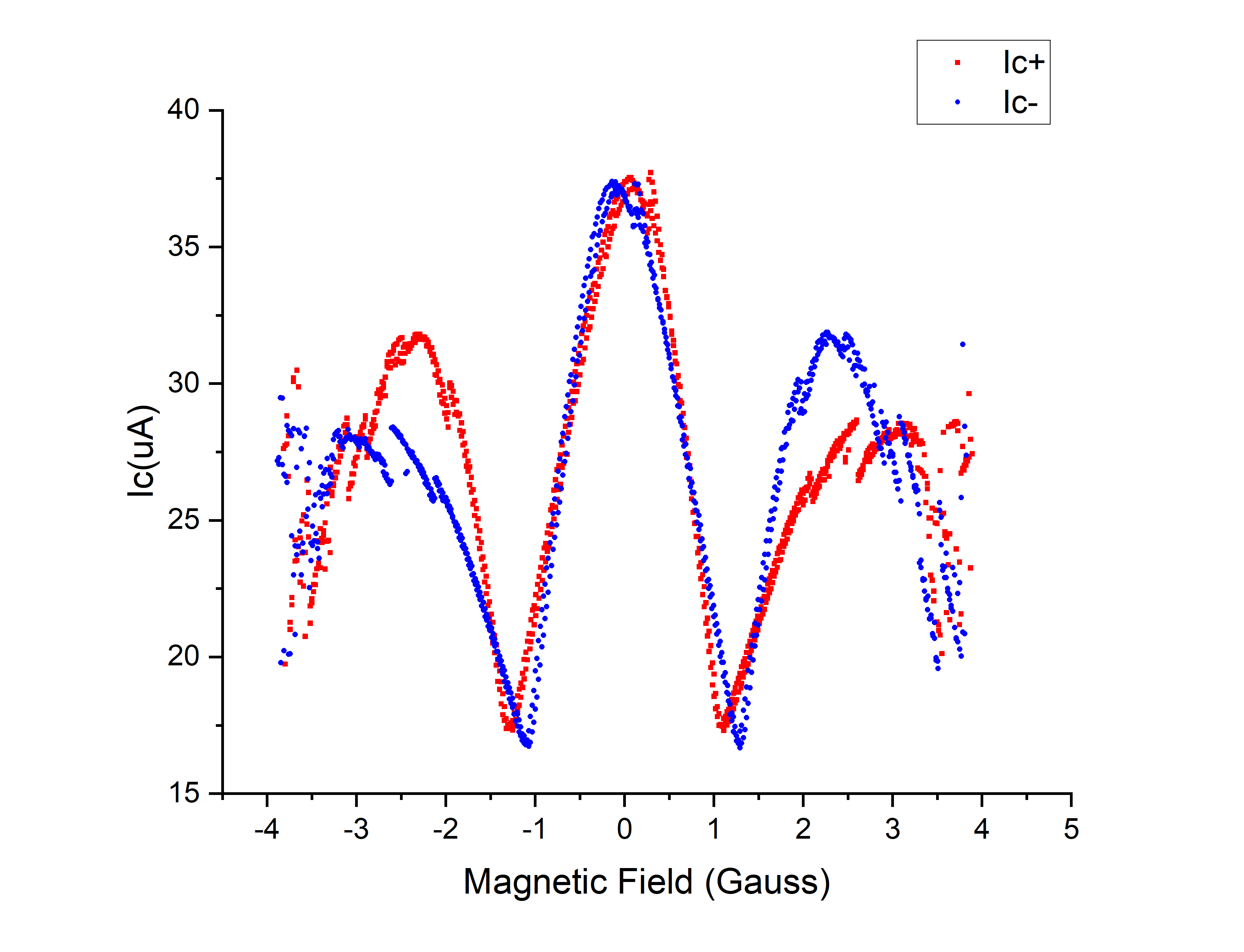}
    \caption{}
    \label{fig:device_c_IB_raw}
  \end{subfigure}

  \vspace{0.3cm}

  \begin{subfigure}[t]{1\linewidth}
    \centering
    \includegraphics[width=\linewidth]{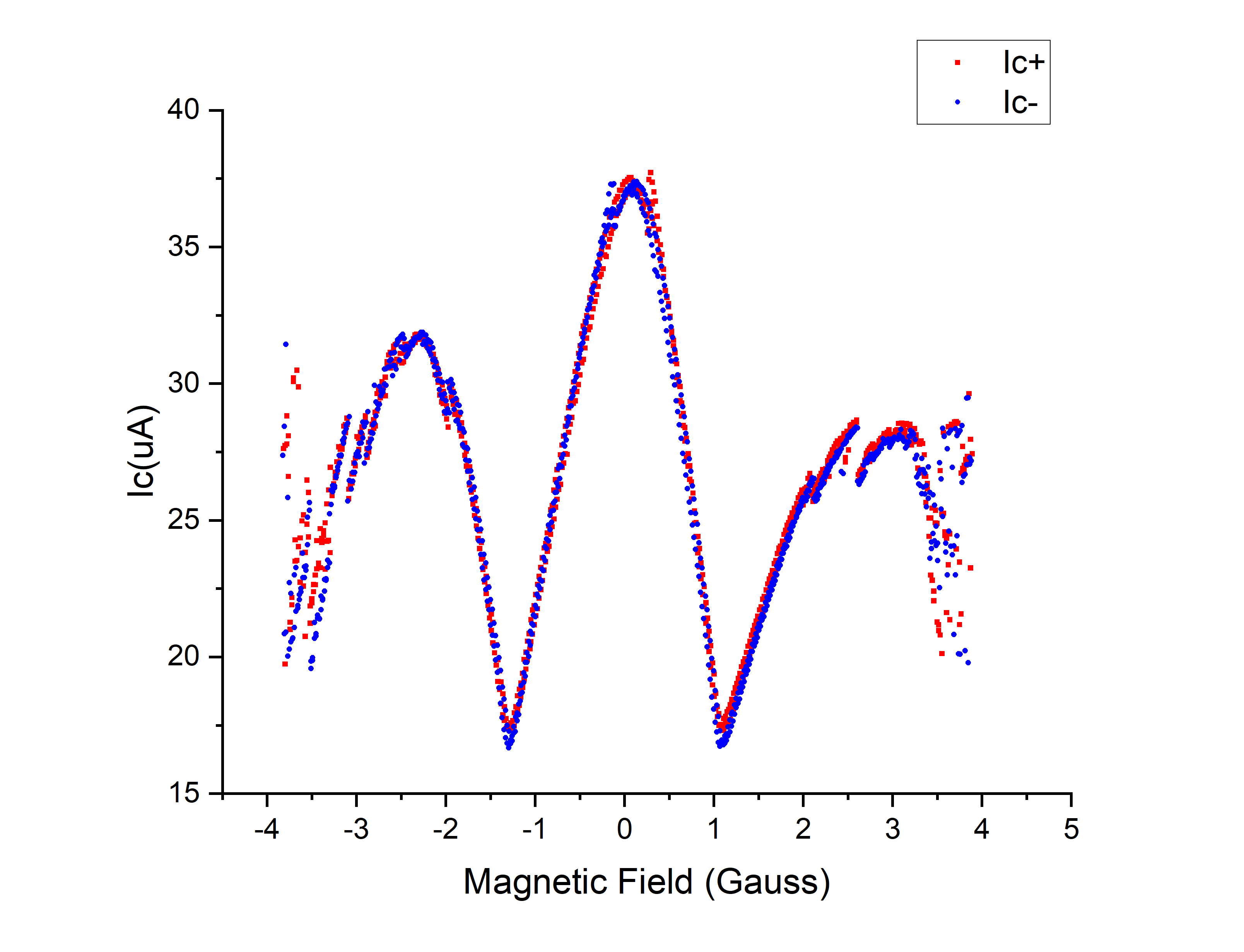}
    \caption{}
    \label{fig:device_c_IB_symmetry}
  \end{subfigure}

  \caption{Magnetic-field dependence of the switching-current envelopes for Device C. 
  (a) Comparison between the positive switching current $I_{c,+}(B)$ and the magnitude of the negative switching current, $-I_{c,-}(B)$. 
  (b) Comparison between $I_{c,+}(B)$ and the field-reversed negative branch, $-I_{c,-}(-B)$, demonstrating that Device C approximately preserves $IB$ symmetry.}
  \label{fig:device_c_IB}
\end{figure}

Device C, whose SEM image is shown in Fig.~\ref{fig:sem}b, is an Al-coated Ag nanowire device and also preserves \(IB\) symmetry. 
The corresponding switching-current data are shown in Fig.~\ref{fig:device_c_IB_symmetry}. 
Compared with Devices A and B, Device C exhibits a distinct modulation pattern with a broader central peak and sharper side minima over a substantially wider magnetic-field range, extending to \(|B|<3.5~\mathrm{G}\).

The switching-current envelope is not completely smooth; instead, it contains several abrupt changes and discontinuity-like features, particularly near the side lobes. 
Nevertheless, these features are reproduced after the transformation $I_{c,-}(B) \rightarrow -I_{c,-}(-B)$, and the two branches remain in close agreement. 
This demonstrates that the $IB$ symmetry is maintained even in the presence of a wider modulation envelope and local breaking-point-like features, indicating that symmetry preservation is robust across different device geometries and templating approaches.


\subsection{Breaking of symmetry}

\begin{figure}[htbp]
  \centering
  \includegraphics[width=1\linewidth]{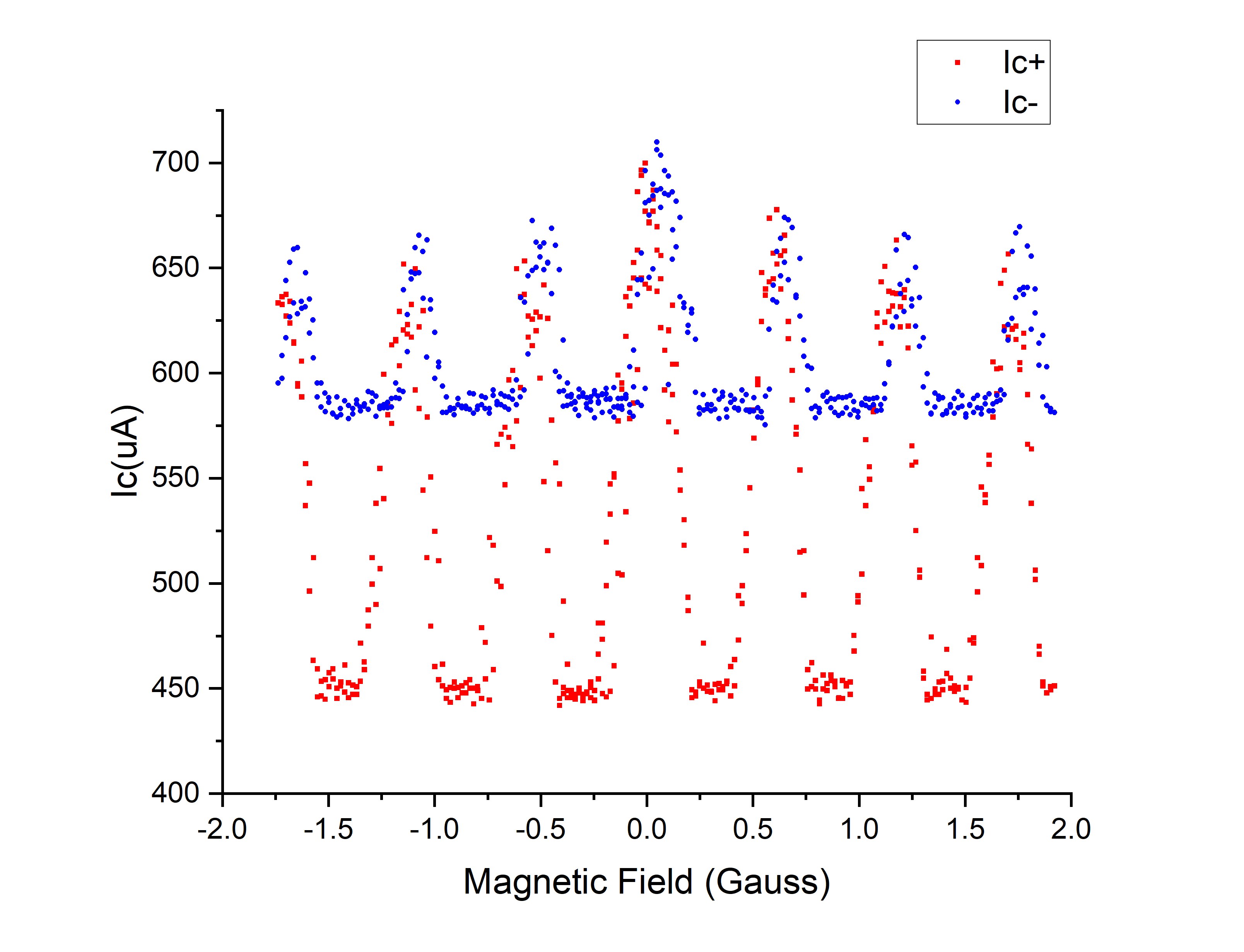}
  \caption{Magnetic-field dependence of the critical current for Device D, which is a Sn-based SIS junction. We explain the symmetry breaking by assuming metallic weak links between and top and bottom films of the junction.}
  \label{fig:deviced_I_sym_breaking}
\end{figure}

\begin{figure}[htbp]
  \centering
  \includegraphics[width=1\linewidth]{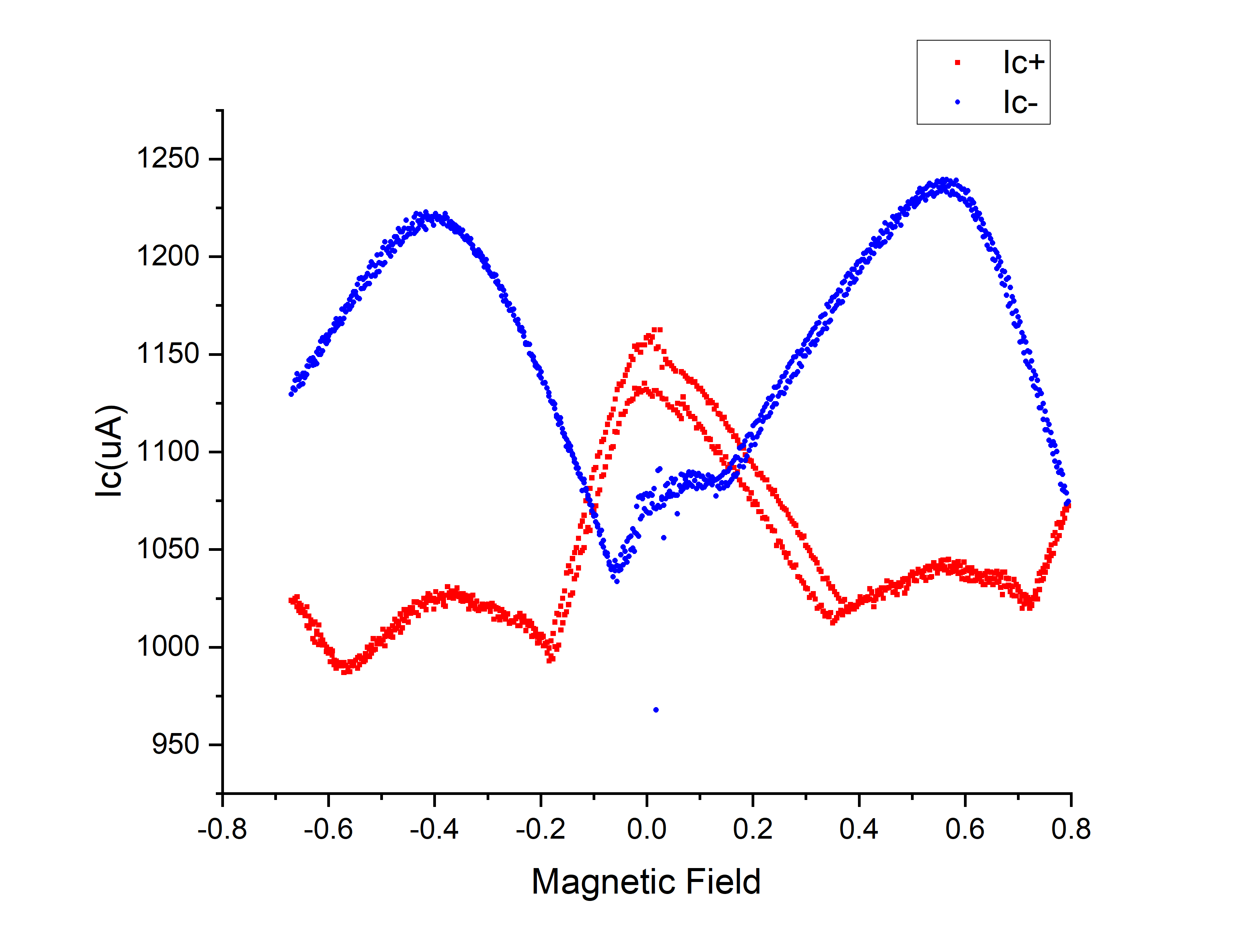}
  \caption{Magnetic-field dependence of the critical current for a device similar to Device D. It is primarily an SIS tin junction, but it has some metallic links between the plates of the junction.}
  \label{fig:devicedb_I_sym_breaking}
\end{figure}

\begin{figure}[htbp]
  \centering
  \includegraphics[width=1\linewidth]{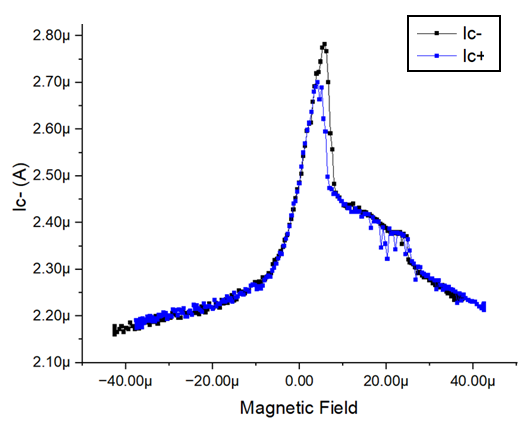}
  \caption{$I_c(B)$ characteristics of Device E, an array of Nb square islands periodically positioned on a topological insulator (BiSeTe) surface.}
  \label{fig:device_e}
\end{figure}

In contrast to the symmetry-preserved behavior of Devices A--C shown in Figs.~\ref{fig:device_a_IB}--\ref{fig:device_c_IB}, we also observe clear violations of \(IB\) symmetry. 
These symmetry-breaking behaviors occur in hybrid junctions involving SIS channels and metallic bridges in parallel, as shown for Device D in Figs.~\ref{fig:deviced_I_sym_breaking} and \ref{fig:devicedb_I_sym_breaking}, and also in the superconductor--topological-insulator--superconductor array shown in Fig.~\ref{fig:device_e}.

In contrast to the symmetry-preserved behavior described above, we also observe clear violations of IB symmetry. These symmetry-breaking behavior is observed on hybrid junctions which involve SIS junctions and metallic bridges in parallel (Device D), and also on superconductor-topological insulator-superconductor (S-TI-S) junction arrays.  In both cases, the positive and negative switching-current envelopes, $I_{c,+}(B)$ and $I_{c,-}(B)$, were measured under identical current-ramp conditions. Then $I_{c,+}(B)$ was compared to the transformed negative branch $-I_{c,-}(-B)$. For a perfectly IB-symmetric device, these two curves coincide, or equivalently follows Eq.~\ref{eq:ib_symmetry}
Instead, the hybrid devices exhibit systematic deviations from this relation, presenting the instances of breaking the simple $(I,B)\rightarrow(-I,-B)$ inversion symmetry of the critical-current function.

The first category of $IB$-symmetry breaking is illustrated in Fig.~\ref{fig:deviced_I_sym_breaking}. 
In this case, the $I_{c,+}(B)$ and the transformed branch $-I_{c,-}(-B)$ appear to roughly coinside on the peaks. Yet, the levels of the positive and negative critical currents are very different in the regions between the peaks. Note also that the peak positions are about the same for both polarities.
 
The main extrema and abrupt switching features occur at approximately the same magnetic-field values for the two branches, indicating that the field-dependent modulation pattern is largely preserved. 
However, at a given magnetic field, the two branches can exhibit substantially different critical-current values. 
This difference is especially evident near the flat-bottom minima, where one branch remains pinned near a lower switching-current plateau while the other branch recovers to a higher value. 
Thus, although the overall envelope shape and characteristic field scales remain qualitatively similar, the two current polarities are not related by the expected simultaneous inversion of current and magnetic field.

The second category of \(IB\)-symmetry breaking is shown in Fig.~\ref{fig:devicedb_I_sym_breaking}. 
This behavior is stronger than the peak-location-preserving case in Fig.~\ref{fig:deviced_I_sym_breaking}. 
After applying the \(IB\)-symmetry transformation, the positive branch \(I_{c,+}(B)\) and the transformed negative branch \(-I_{c,-}(-B)\) still do not collapse onto a common envelope. 
The discrepancy is not limited to a vertical difference in switching-current magnitude. 
Instead, the two branches show both a horizontal displacement in magnetic field and a change in envelope shape. 
The positions of the dominant maxima are different for the two current polarities, and the relative importance of the central and side features appears to be exchanged. 
In addition, the local slopes and plateau-like regions are different, indicating that the two current directions do not simply follow the same interference pattern with a small field offset.

This type of symmetry breaking is important because it cannot be explained by a trivial calibration error or by a rigid magnetic-field shift. 
A field offset would move the entire curve along the \(B\)-axis while preserving the relative spacing, peak hierarchy, and local envelope shape. 
In Fig.~\ref{fig:devicedb_I_sym_breaking}, however, the transformed branches have different shapes and different dominant features. 
This suggests that the positive and negative current ramps access different sequences of metastable vorticity states in the hybrid weak-link network. 
Within the hybrid SIS--metallic-link model, the coexistence of sinusoidal SIS channels and approximately linear metallic-link channels creates a multivalued energy landscape. 
Different current polarities can then preferentially select different switching branches, causing the observed mismatch in both peak position and peak height.

Thus, Fig.~\ref{fig:devicedb_I_sym_breaking} represents a shape-distortion-type \(IB\)-symmetry breaking. 
In this regime, the device does not merely break current-inversion symmetry at fixed field; it also fails the stronger combined symmetry test \(I_{c,+}(B)=-I_{c,-}(-B)\). 
This provides evidence that the symmetry breaking is tied to polarity-dependent branch selection in a hybrid weak-link energy landscape, rather than to a conventional single-valued SIS interference pattern.

Taken together, these measurements show that IB symmetry breaking in hybrid SIS+metallic bridges devices can occur in at least two distinct forms: (i) a peak-location-preserving symmetry breaking, which appears to obey B-symmetry, but not I-symmetry, and (ii) a \emph{shape-distortion-type} asymmetry, in which the locations of primary and secondary maxima swap their magnetic field values. The coexistence of these two behaviors suggests that the symmetry breaking is not caused solely by a trivial residual magnetic-field offset, but instead reflects polarity-dependent physics introduced by the hybrid weak-link structure and wider electrodes.

Another example is given by Device E, based on a topological insulator film. The results are shown in Fig.~\ref{fig:device_e}. Generally speaking, this device exhibits a superconducting diode effect, in which the critical current maximum depends on the direction of current flow, so that the positive and negative switching currents are no longer equal in magnitude\cite{song-2023}. Also, the device obeys IB-symmetry approximately. In that respect it is similar to Ref.\cite{zvzr-flw6}.
If the IB-symmetry transformation is applied, the positive and negative switching-current branches largely follow the same B-field dependence (Fig.~\ref{fig:device_e}) over most of the measured magnetic field range. 
However, a clear symmetry breaking appears near the central maximum: the peak position and peak magnitude of $I_{c,+}(B)$ and $-I_{c,-}(B)$ are slightly offset from one another. 
Away from this central region, the two branches nearly overlap, indicating that the asymmetry is localized primarily around the maximum critical-current region rather than distributed across the entire interference pattern. In what follows we will consider the justification for the IB-symmetry and discuss possible reasons for its violation observed in some samples.

\section{Model}

To study understand the $IB$ symmetry, we model different classes of superconducting quantum interference devices (SQUID) with various types of Josephson junctions (JJ), including nanowires characterized by linear current-phase relationship (CPR), extended superconductor-insulator-superconductor (SIS) junctions characterized by sinusoidal CPR, and combinations of those.  In the case of SIS junctions, the critical current is a single-valued function of the applied magnetic field. This function obeys I-symmetry and B-symmetry and IB-symmetry, as will be discussed in the following sections. For devices containing nanowires the CPR is a multi-valued function since the device supports different possible vorticity sates. The critical current function of the magnetic field obeys IB-symmetry in such devices, but it is not I-symmetrical so the devices act as superconducting diodes.

\subsection{Multiple-nanowire SQUID model}

Here we analyze SQUIDs containing many parallel superconducting
nanowires connected between large superconducting electrodes. For each
nanowire, we use an approximately linear current--phase relation\cite{hopkins_SQUID,pekker-2005,model1},
\begin{equation}
    I_i(\phi_i)=\frac{I_{c,i}}{\phi_{c,i}}\phi_i ,
    \qquad |\phi_i|\leq \phi_{c,i},
\end{equation}
where $I_i$ is the supercurrent through the $i$th nanowire, $I_{c,i}$ is the
critical current of the $i$th nanowire, $\phi_i$ is the superconducting phase
difference across that nanowire, and $\phi_{c,i}$ is its critical phase for each nanowire. All these variables represent arrays of real numbers and the number of elements in each array is $n$, which is the total number of nanowires in the considered SQUID. In
this model, a switching event of the SQUID (and the critical current of the entire SQUID) occurs when at least one nanowire reaches its
critical phase, and correspondingly, its critical current. If one of the wires hits its critical current the vorticity distribution of the device must change. Therefore we consider so-called vorticity stability regions (VSR)\cite{model1}, the boundaries of the superconducting regions in
the current--magnetic field $(I,B)$ plane correspond to the critical-phase
condition of one or more nanowires, for a given vorticity distribution (numbers of vortices in the SQUID loops). 

\begin{figure*}[t]
    \includegraphics[width=0.32\linewidth]{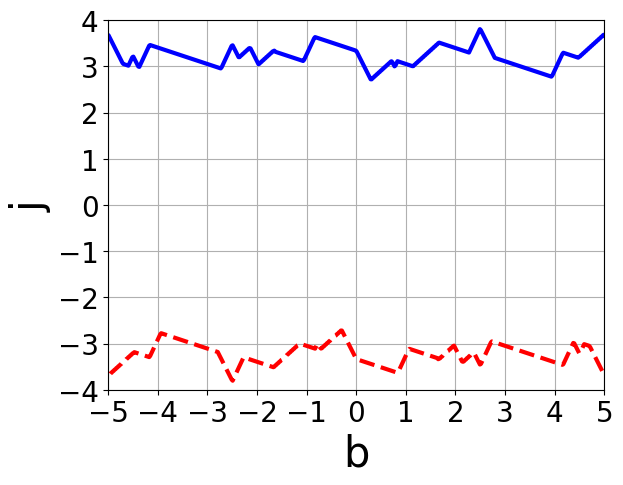}
    \includegraphics[width=0.32\linewidth]{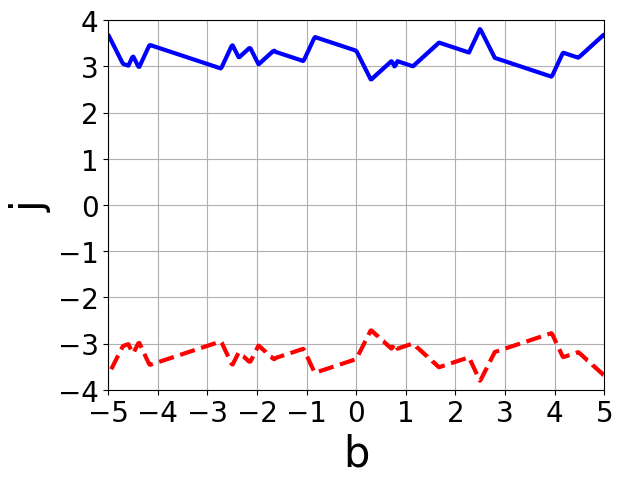}
    \includegraphics[width=0.32\linewidth]{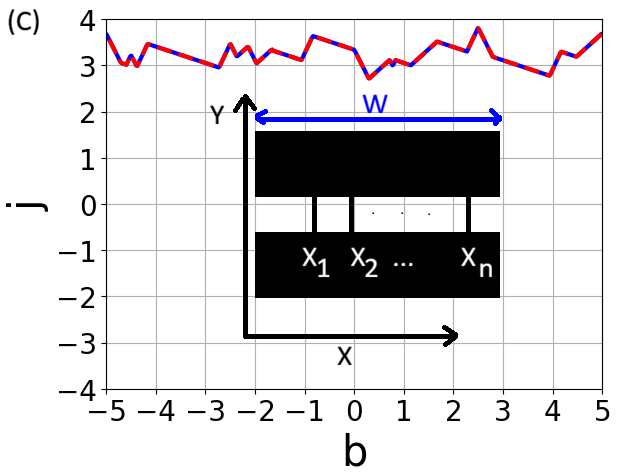}
    
    \caption{$IB$ symmetry in $I_c$(b) curve for four nanowires of positions $x_i=[0, 0.3, 0.6, 1]$, critical phases of $\phi_{c,i}=[4\pi, 2\pi, \pi, 3\pi]$, and critical currents of $I_{c,i}=[1.2, 0.7, 1.3, 0.8]$. The normalized positions are defined as $x_i=(X_i-X_1)/(X_n-X_1)$. The normalized total current is $j=(<I_{c,i}>)^{-1}\sum I_i$, where $\sum I_i$ is the total current through the multi-wire SQUID and $<I_{c,i}>$ is the average critical current. The normalized magnetic field is $b=B/\Delta B$. (a) $I_{c,+}(b)$ is shown as the blue solid curve, and $I_{c,-}(b)$ is shown as the red dashed curve. (b) $I_{c,+}$ is unchanged, while $B$ is inverted for the negative critical current so that the dashed curve is $I_{c,-}(-b)$. (c) $I_{c,+}(b)$ is shown unchanged, while the dashed curve represents the function $-I_{c,-}(-B)$, i.e. the negative critical current, with inverted magnetic field and multiplied by -1. Thus, $-I_{c,-}(-b)$ coincides with $I_{c,+}(b)$. This illustrates the $IB$ symmetry of the critical current. The insert shows a schematic of the multiple-wire SQUID with the coordinates indicated. The black color represents a superconducting material. The magnetic field (not shown) is perpendicular to the plane of the device. The width of the electrodes is $W$. }
    \label{fig:4_wire_icb_symmetry}
\end{figure*}

\begin{figure}
    \centering
    \includegraphics[width=1\linewidth]{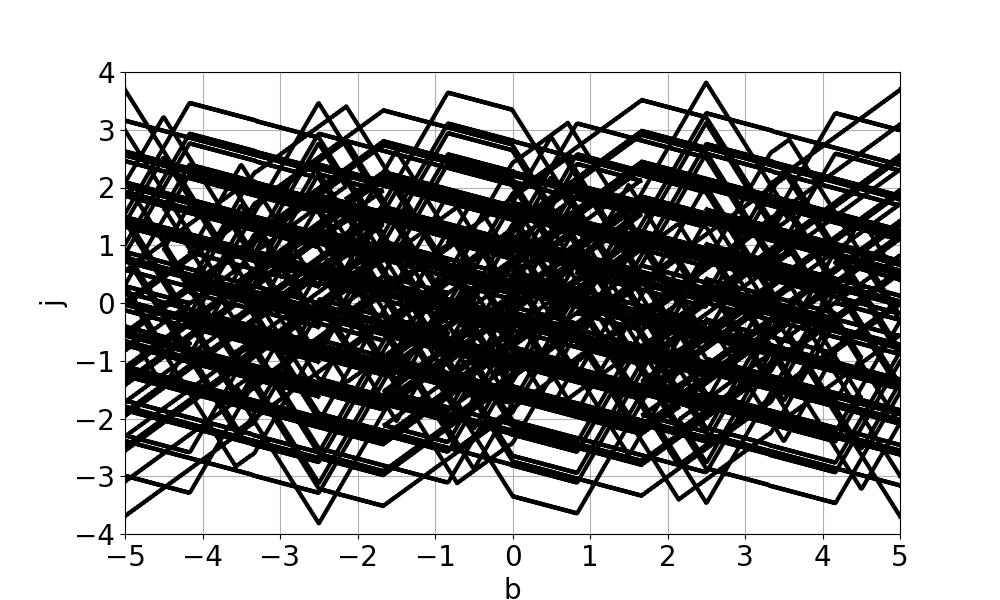}
    \caption{All Vorticity Stability Regions of a disordered 4-wire SQUID presented in Fig.\ref{fig:4_wire_icb_symmetry}. Here, the positions of the nanowires are $[0, 0.3, 0.6, 1]$, critical phases of $[4\pi, 2\pi, \pi, 3\pi]$, and critical currents of $[1.2, 0.7, 1.3, 0.8]$. The top and the bottom envelope curves correspond to those of Fig.~\ref{fig:4_wire_icb_symmetry}.  }
    \label{fig:placeholder}
\end{figure}

Because all nanowires are connected by the same superconducting electrodes,
the phase differences across different nanowires are not
independent. They are related by the phase-correlation condition around closed
superconducting loops formed by pairs of weak links and the electrodes. For the
$i$th and $j$th wire, this relation can be written as
\begin{equation}\label{eq:PCE}
    \phi_j
    =
    \phi_i
    +2\pi \frac{(X_j-X_i)B}{(X_n-X_1)\Delta B}
    -2\pi v_{ij}
\end{equation}
We refer to this formula as Meissner phase equation (MPE). Here, $\phi_i$ and $\phi_j$ are the superconducting phase differences across
the $i$th and $j$th weak links, respectively. This formula represents a generalized Little-Parks effect. The integer $v_{ij}$ counts the
net number of fluxoids, or vortices, enclosed between these two weak links.
The quantity $\Delta B_1$ is the magnetic-field normalization constant
corresponding to the magnetic field period of the SQUID containing just the first wire at position $X_1$ and the last wire at position $X_n$. As was explained in
Refs.~\cite{hopkins_SQUID,pekker-2005}, it is expressed as
\begin{equation}
    \Delta B
    =
    \frac{\pi^2}{8G}\frac{\Phi_0}{Wd},
\end{equation}
where $\Phi_0=h/2e$ is the superconducting flux quantum, $d=X_n-X_1$ is the
distance between the first and last nanowire, $W$ is the width of the
electrodes, and $G=0.916$ is Catalan's constant. With this definitions, the
magnetic phase shift between two weak links separated by a distance
$(X_j-X_i)$ is
\[
    2\pi \frac{(X_j-X_i)}{(X_n-X_1)}\frac{B}{\Delta B_1}.
\]
Thus, in the limit of a two-nanowire SQUID in which $X_i=X_1$ and $X_j=X_n=X_2$, the phase
shift between the two wires is $2\pi$ if $B=\Delta B_1$. Note that, according to the Little-Parks effect, this phase shift can be reduced to zero if one vortex enters in the space between the nanowires, i.e., if $v_{ij}=v_{12}=1$, in agreement with Eq.\ref{eq:PCE}.

Equation~\ref{eq:PCE} follows from phase winding around a closed
superconducting loop. It is the total phase winding around the loop that is
quantized in integer multiples of $2\pi$; the local phase difference $\phi_i$
across an individual weak link is a continuous phase variable and is not itself
restricted to integer multiples of $2\pi$. The magnetic-field term represents
the phase-winding contribution accumulated in the electrodes, between the two nanowires . As in
Refs.~\cite{hopkins_SQUID,pekker-2005,model1}, we assume that nanowire is
short in the current-flow direction, so the contribution of the vector potential
along the length of each nanowire is neglected, due to the assumption that the wire is sufficiently short. Under this
approximation, the phase bias correlation between the nanowires is primarily dictated by the phase winding in the electrodes, through the Meissner current effect, which generates this phase winding.

For a multi-nanowire SQUID (MW-SQUID), the allowed superconducting states form closed
domains in the current--magnetic field plane. These domains are called
vorticity stability regions (VSRs)~\cite{model1}. An example of the VSR diagram is shown in Fig.\ref{fig:placeholder}. Each VSR corresponds to a
particular vorticity configuration, and its boundary is determined by the
condition that at least one nanowire reaches its critical phase (critical current). The maximum
positive and negative critical-current branches are obtained from the upper and
lower envelopes of all accessible VSRs, as shown in Fig.\ref{fig:4_wire_icb_symmetry}. But in some cases the experimental results reflect boundaries of multiple VSR, producing a multi-valued function of the critical current versus magnetic field, corresponding to various values of the vorticity, as, for example, in Fig.\ref{fig:device_b_IB}. This happens because once the boundary of a VSR is reached, a phase slip must occur to allow vortices to jump between the cells or to allow new vortices to enter or exit. Such phase slips, representing a vortex motion, generate heat and may be able to switch the device to its normal (resistive) state.

The model outlined above predicts that the multiple-nanowire SQUID devices must obey the $IB$ symmetry, as shown in Fig.\ref{fig:4_wire_icb_symmetry}. When all vorticity states are included, each VSR has a
partner under simultaneous inversion of current, magnetic field, and vorticity,
\begin{equation}
    (I,B,V)\rightarrow(-I,-B,-V),
\end{equation}
where $V$ denotes the vorticity array. This state-resolved symmetry is referred
to as $IBV$ symmetry. Taking the envelope over all accessible vorticity states
then gives the experimentally relevant $IB$ symmetry of the critical current,
\begin{equation}
    I_{c,+}(B)= -I_{c,-}(-B).
\end{equation}
It is applicable if the vorticity is always optimized so that the maximum possible critical current is detected. It is also applicable if the vorticity is different in each measurement but the distribution of the realized vorticity states is the same in magnitude for both positive bias current and the negative bias current. This result also provides a hin of the possible breaking of the IB-symmetry. The IB-symmetry can be broken if the vorticity sate distribution is different not only in sign but in magnitude for the bias currents of opposite polarities. 

The similar argument also shows how $IB$ symmetry can be broken if the vorticity
configuration is constrained, for example because a vortex is trapped or pinned
in the electrodes of the device. Then the partner state required by
$V\rightarrow -V$ may no longer be accessible. In this case, the full $IBV$
symmetry of the VSR set is not reflected in the measured critical-current
envelope, and the observed $IB$ symmetry can be violated.

\begin{figure}
    \centering
    \includegraphics[width=1\linewidth]{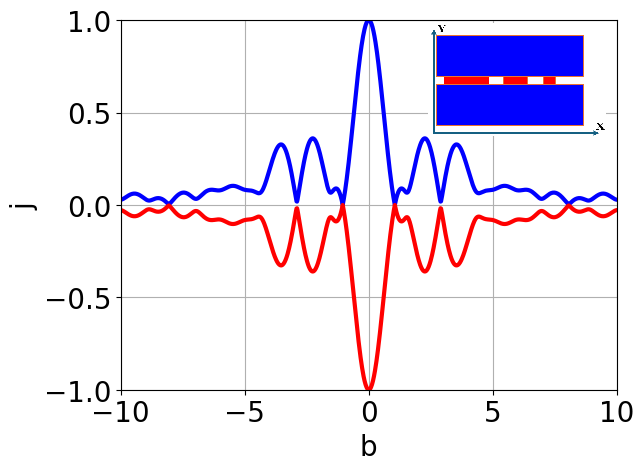}
    \caption{A disordered SIS is illustrated in the insert. Blue regions are superconducting films and the red regions represent segments where the critical current is larger than zero. The dimensions of these regions along the x-axis can be arbitrary in this model. The current density is proportional to the phase difference, i.e., the CPR is assumed sinusoidal. This is because the red region as presumed a thin insulator or normal metal, so that the junctions can be either SIS or SNS in this model. The white color represents regions where the critical current density is zero. Main graph: Normalized critical current of such multiple-extended-SIS-junction SQUID is plotted versus normalized magnetic field. The model predicts that the junction array always obeys the I-symmetry and the B-symmetry. Here, extended SIS junctions are located at the following normalized positions: 0 to 0.2, 0.32 to 0.73, and 0.86 to 1. There are a total of three extended junctions. The critical current is assumed zero between the junctions, i.e., in the interval $0.2<x<0.32$ and $0.73<x<0.86$.}
    \label{fig:jj_symmetry}
\end{figure}

\subsection{Model with multiple extended superconductor-insulator-superconductor junctions}

Here, we present a model of a SQUID made with many SIS extended JJs. Similarly to the previous section, the phase difference is set by the magnetic field, the dimensions of the electrodes\cite{hopkins_SQUID,pekker-2005}, and the position along the electrodes, $X$. The extended-SIS-junction SQUID device is illustrated in Fig.\ref{fig:jj_symmetry}(insert), where the electrodes are blue, the regions with zero critical current density are white and the segments with a finite critical current density, $J_c(X)>0$, are red. Here $X$ is the horizontal position along the extended junction, shown in Fig.\ref{fig:jj_symmetry}(insert). This multi-junction array is described by a local current phase relationship: 

\begin{equation}\label{eq:SIS_CPR}
    J(X) = J_c(X)\sin(\phi(X))
\end{equation}
where $J(X)$ is the local current density, $J_c(X)$ is the position-dependent critical current density and $\phi(X)$ is the local phase difference between the superconducting electrodes. Note that with these definitions $J_c(X)$ is zero in the white colored regions and it is larger than zero in the the red-colored regions of Fig.\ref{fig:jj_symmetry}(insert).

In what follows we used a normalized position along the electrodes, defined a $x=(X-X_L)/(X_R-X_L)$, where $X_L$ is the left end of the left-most SIS junction and $X_R$ is the coordinate of the right end of the right-most junction. In other words, $J_c=0$ for $X<X_L$ and for $X>X_R$.

We also introduce a normalized magnetic field as $b=B/\Delta B$. The standard SQUID period $\Delta B=(\pi^2)/(8G)(\Phi_0/Wd)$, where G=0.916, $W$ is the width of the electrodes shown in Fig.\ref{fig:jj_symmetry}(insert) and $d=X_R-X_L$. In this model the magnetic field focusing effect is neglected, assuming that the electrodes are made of a thin superconducting film, so that the effective magnetic field penetration depth is larger than the size of the electrodes\cite{hopkins_SQUID,pekker-2005}. 

The Meissner phase correlation equation, analogous to Eq.~\ref{eq:PCE}, is generalized to a continuous form as follows 

\begin{equation}
    \phi(x,b) = \phi_1 + 2\pi b x
\end{equation}

Here, $b = B/\Delta B$ and $x = (X - X_L)/(X_R - X_L)$ Here, $X$ is the physical coordinate along the lateral direction of the junction array.
The coordinates $X_L$ and $X_R$ define the left and right boundaries of the modeled array region, respectively.
For the extended SIS-junction model used here, $X_L$ is taken as the left edge of the first junction segment, while $X_n$ is the right edge of the last junction segment. With this convention, (x=0) corresponds to the left boundary of the array and (x=1) corresponds to the right boundary. 

The total current of the device $I_{tot}(b)$ is the integral of Eq.~\ref{eq:SIS_CPR} from $x=0$ to $x=1$:

\begin{equation}
    I_{tot}(b) = \int_0^1 J_c(x)\sin(\phi(x,b))dx
\end{equation}
Interestingly, the total current as a function of $b$ resembles a Fourier transform of the critical current density function $J_c(x)$. Note also that the vorticity plays no role in this model and the total current is a single-valued function of the magnetic field.  This is because the CPR is sinusoidal and $\sin (\phi+2\pi v)=\sin(\phi)$ for any integer vorticity number $v$.

An example result of this model is given in Fig.~\ref{fig:jj_symmetry}. The assumed critical current density, $J_c(x)$, is a piecewise function defined as $J_c(x)=1$ if $x$ is in the ranges of: $0<x<0.2$, $0.32<x<0.73$, and $0.86<x<1$, and $J_c(x)=0$ otherwise. 
In this numerical tests (and all others), we found that the resulting critical current curve is always individually current-inversion symmetric (I-symmetry) and magnetic field inversion symmetric (B-symmetry). Therefore it is also  IB-symmetric, as illustrated in Fig.~\ref{fig:jj_symmetry}. In other words, critical current curves for randomly distributed and extended SIS JJ arrays are always symmetric with respect to current inversion and magnetic field inversion.  

% \begin{figure}
%     \centering
%     \includegraphics[width=\linewidth]{imgs/SIS_exp_data.png}
%     \includegraphics[width=\linewidth]{imgs/best_fit_3_SIS.png}
%     \caption{(Left) Experimental critical current data of an many SIS extended and random junction device. (Right) Best fit to experimental critical current data with model.}
%     \label{fig:exp_fitting}
% \end{figure}

% \begin{figure}
%     \centering
%     \includegraphics[width=1\linewidth]{imgs/exp_fit.png}
%     \caption{Experimental critical current data (red) plotted against the best 3 SIS junction fit (blue) of $[0, 0.06, 0.13, 0.77, 0.99, 1]$. }
%     \label{fig:placeholder}
% \end{figure}

% \begin{figure}
%         \centering
%         \includegraphics[width=1\linewidth]{imgs/exp_fitting.png}
%         \caption{Experimental critical current data (red) plotted against the best 3 SIS junction fit (blue) of $[0, 0.06, 0.13, 0.77, 0.99, 1]$. }
%         \label{fig:placeholder}
%     \end{figure}

This symmetry is different from the symmetry found in the nanowire-SQUID model.
For arrays of SIS junctions with sinusoidal current--phase relations,
the critical-current pattern is individually symmetric under current
inversion and magnetic-field inversion. In contrast, the multiple-nanowire
SQUID is generally protected only under the combined $IB$ transformation, after
the full set of vorticity states is considered. Thus, arrays of nanowires can be used as superconducting diodes and range controllers, while SIS junction arrays cannot in this simple realization. Such symmetry difference is somewhat unexpected since the individual CPR for each nanowire is assumed an antisymmetric function, same as for SIS junctions. Note also that these results for SIS junctions would also apply to SNS junctions, such as those presented in Fig.\ref{fig:device_e}

\subsection{Hybrid SIS--Dayem-Bridge Model}
\label{subsec:hybrid_jj_nanowire_modeling}

Before introducing the hybrid model, it is useful to distinguish the observed symmetry breaking (Fig.~\ref{fig:deviced_I_sym_breaking}) from the extended random junction discussed in the previous section. In the previous section we have found, through our simple model, that the I-symmetry and the B-symmetry are preserved if the critical current density is modulated randomly but the local CPR remains sinusoidal. Yet, our Sn/SnO\(_x\)/Sn junctions (Device D), shown in Fig.~\ref{fig:deviced_I_sym_breaking}, exhibits a clear I-symmetry breaking while B-symmetry is preserved, at least approximately.

We therefore model Device D as an effective hybrid weak-link array in which sinusoidal SIS channels coexist with metallic-link (ML) channels having linear CPRs, analogous to superconducting nanowires. Note that Such systems are interesting because they can be used to create superconducting rectifiers, as have been discussed in Ref.\cite{zvzr-flw6}.

Our present model is not intended to be a literal geometrical reconstruction of the Sn/SnO\(_x\)/Sn sandwich junction. 
Instead, it represents the junction as a small number of effective parallel current-carrying channels. 
Some channels are modeled as metallic-link or Dayem-bridge-like channels with linear CPRs, while the others are modeled as SIS channels with sinusoidal CPRs. 
This is physically motivated by the fact that the device contains Sn/SnO\(_x\)/Sn tunnel-junction regions in parallel with possible metallic weak links caused by imperfections of the oxide layer.

\begin{figure} [htbp]
    \centering
    \includegraphics[width=0.5\linewidth]{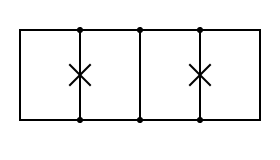}
    \caption{Schematic of the five-channel hybrid model. The channels are ordered ML--SIS--ML--SIS--ML. The middle ML channel, \(k=2\), is used as the reference channel, \(\phi_{\mathrm{ref}}=\phi_2\). The two SIS channels are placed symmetrically about this middle channel, while the outer ML channels provide additional vorticity-dependent switching levels.}
    \label{fig:schematic}
\end{figure}

The minimal model used here is a five-channel weak-link array with normalized channel positions
\begin{equation}
    x_k = [0,0.25,0.5,0.75,1.0],
    \qquad k=0,1,2,3,4,
    \label{eq:hybrid_positions}
\end{equation}
in a ML--SIS--ML--SIS--ML sequence. The effective five-channel geometry is summarized schematically in Fig.~\ref{fig:schematic}. 
The model is not intended to reproduce the microscopic Sn morphology exactly; instead, it captures the minimal ingredients needed for the observed symmetry breaking: two sinusoidal SIS-like channels and three metallic-link channels with approximately linear current--phase relations. 
The middle ML channel is chosen as the reference channel, while the two SIS channels are placed symmetrically about it. Here, ML denotes a metallic-link or Dayem-bridge-like channel with a linear CPR and SIS denotes a superconducting tunnel-junction channel with a sinusoidal CPR. 

The ML channels are \(k=0,2,4\) and the SIS channels are \(k=1,3\).
For the hybrid model we use
\begin{equation}
    I_k(\phi_k)=
    \begin{cases}
        K_k\phi_k, & k=0,2,4,\\
        I_{J,k}\sin\phi_k, & k=1,3.
    \end{cases}
    \label{eq:hybrid_channel_cpr}
\end{equation}
Here \(I_k\) is the current in the channel \(k\), \(K_k\) is the phase stiffness of ML channel \(k\), and \(I_{J,k}\) is the Josephson critical-current of SIS channel \(k\). The total supercurrent is
\begin{equation}
    I_{\mathrm{tot}} = \sum_{k=0}^4 I_k(\phi_k).
    \label{eq:hybrid_total_current}
\end{equation}
The ML channels are stable only while \(|\phi_k|\leq \phi_{c,k}\), where \(\phi_{c,k}\) is the critical phase constant.

We chose the middle ML channel as the reference channel,
\[
    \phi_{\mathrm{ref}} = \phi_2.
\]
The vorticity state is written as
\[
    \mathbf v=(v_0,v_1,v_2,v_3),
\]
where \(v_m\) is the integer vorticity in the cell between channels \(m\) and \(m+1\). We use the same normalized magnetic field as above,
\[
    b = \frac{B}{\Delta B}.
\]
Here \(\Delta B\) is the effective magnetic-field period of the SQUID, as was defined in the previous sections. 

For the five-channel model, the phase relations are
\begin{align}
    \phi_0 &= \phi_{\mathrm{ref}}-\pi b+2\pi(v_0+v_1), \\
    \phi_1 &= \phi_{\mathrm{ref}}-\frac{\pi}{2}b+2\pi v_1, \\
    \phi_2 &= \phi_{\mathrm{ref}}, \\
    \phi_3 &= \phi_{\mathrm{ref}}+\frac{\pi}{2}b-2\pi v_2, \\
    \phi_4 &= \phi_{\mathrm{ref}}+\pi b-2\pi(v_2+v_3).
    \label{eq:hybrid_explicit_phases}
\end{align}

For this hybrid system, two different critical-current definitions can be introduced.

First, one may define a global mathematical maximum and minimum of the total current--phase relation for a fixed magnetic field and vorticity state ("global maximum" criterion):
\begin{align}
    I_{+}^{\mathrm{max}}(b,\mathbf v)
    &=
    \max_{\phi_{\mathrm{ref}}\in \mathcal W(b,\mathbf v)}
    I_{\mathrm{tot}}(\phi_{\mathrm{ref}},b,\mathbf v),
    \label{eq:hybrid_global_positive} \\
    I_{-}^{\mathrm{min}}(b,\mathbf v)
    &=
    \min_{\phi_{\mathrm{ref}}\in \mathcal W(b,\mathbf v)}
    I_{\mathrm{tot}}(\phi_{\mathrm{ref}},b,\mathbf v),
    \label{eq:hybrid_global_negative}
\end{align}
where \(\mathcal W(b,\mathbf v)\) is the allowed phase window (for the free variable $\phi_{\text{ref}}$) in which all ML channels satisfy
\begin{equation}
    |\phi_k|\leq \phi_{c,k},
    \qquad k\in\{0,2,4\}.
    \label{eq:hybrid_validity_window}
\end{equation}

Second, one may define a ML channel-switching criterion ("ML critical phase" criterion). 
In this criterion, switching occurs when one ML channel reaches its critical phase. 
For a switching channel \(k\in\{0,2,4\}\), the switching condition is
\begin{equation}
    \phi_k(\phi_{\mathrm{ref}},b,\mathbf v)=\pm\phi_{c,k}.
    \label{eq:hybrid_branch_switch_condition}
\end{equation}
The corresponding switching current is
\begin{equation}
    I_{\mathrm{sw},\pm}^{(k)}(b,\mathbf v)
    =
    I_{\mathrm{tot}}(\phi_{\mathrm{ref},\pm}^{(k)},b,\mathbf v),
    \label{eq:hybrid_branch_switch_current}
\end{equation}
where \(\phi_{\mathrm{ref},\pm}^{(k)}\) is the value of the reference phase that satisfies Eq.~\ref{eq:hybrid_branch_switch_condition}. 
The remaining ML channels must still satisfy Eq.~\ref{eq:hybrid_validity_window}.

\begin{figure}[htbp]
    \centering
    \includegraphics[width=1\linewidth]{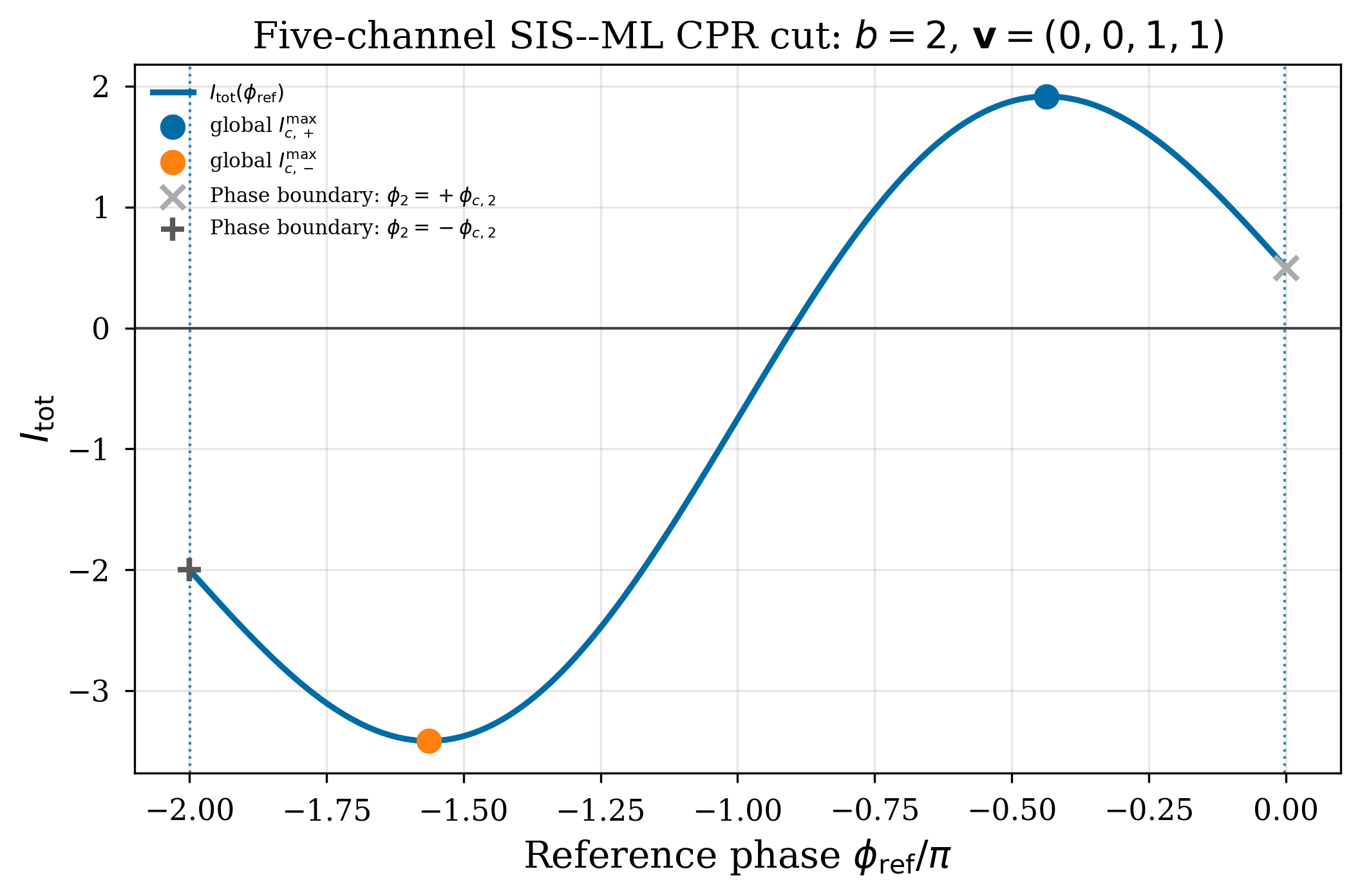}
    \caption{ Comparison of two critical-current definitions in the five-channel SIS--metallic-link model. The curve shows \(I_{\mathrm{tot}}(\phi_{\mathrm{ref}})\) for \(b=2\) and \(\mathbf v=(0,0,1,1)\), over the phase window allowed by the ML critical-phase constraints. The global-extremum definition takes the maximum or minimum of the full CPR. The channel-switching definition instead evaluates the current when an ML channel reaches its critical phase; in this example the marked boundary points correspond to \(\phi_2=\pm\phi_{c,2}\). The two definitions need not coincide. }
    \label{fig:Ic_defs_cpr}
\end{figure}

The distinction between these two criteria is illustrated in Fig.~\ref{fig:Ic_defs_cpr}. The example shows the total CPR of the five-channel hybrid model at \(b=2\) for the vorticity state \(\mathbf v=(0,0,1,1)\). The circles mark the global maximum and minimum of \(I_{\mathrm{tot}}\) over the allowed phase window, while the crosses mark the currents obtained when the middle ML channel reaches \(\phi_2=\pm\phi_{c,2}\). Thus the mathematical maximum identifies the largest current supported by the full CPR, whereas the channel-switching criterion identifies the current at which a particular metallic-link channel becomes normal.

\begin{figure} [htbp]
    \centering
    \includegraphics[width=\linewidth]{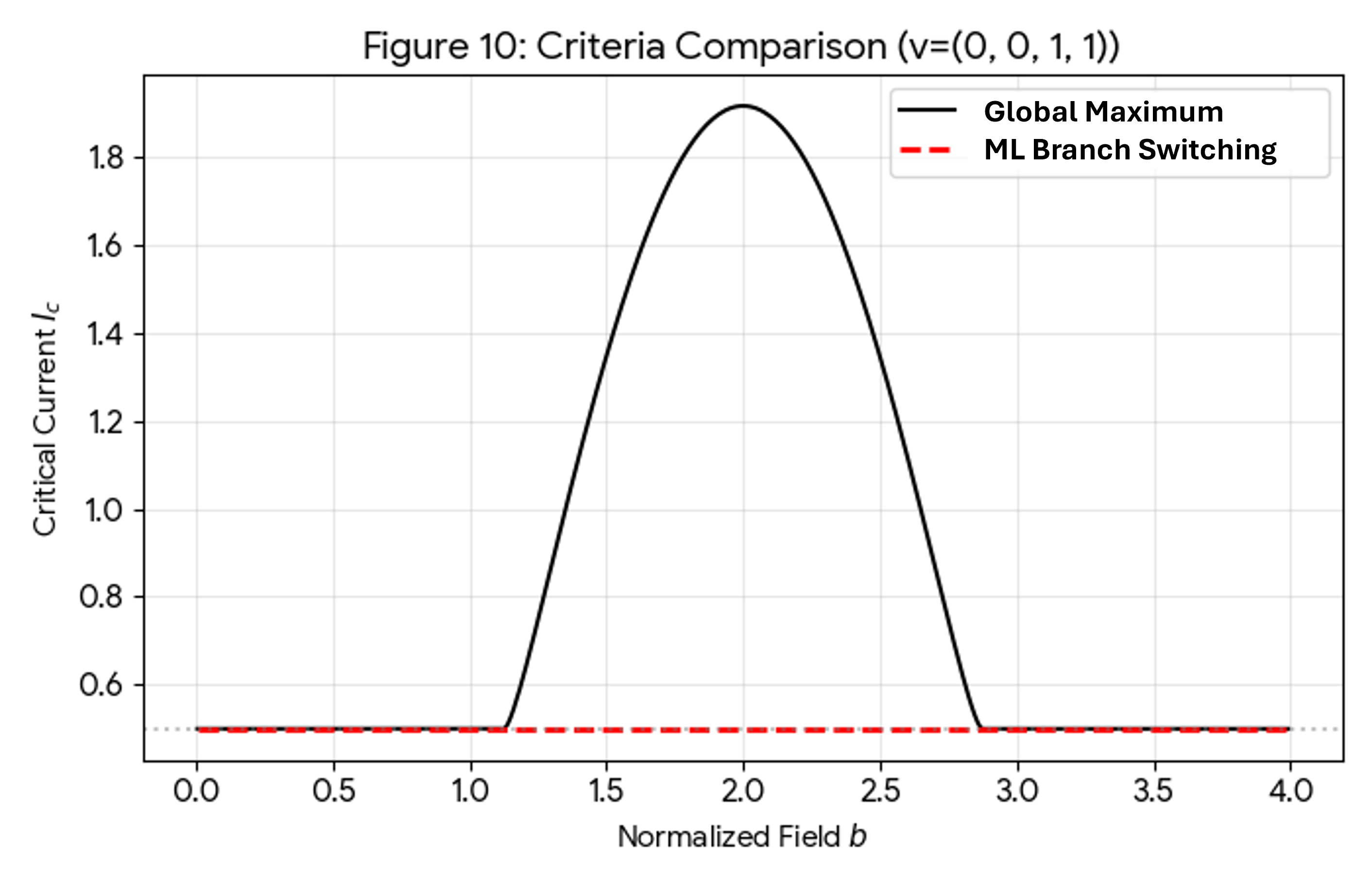}
    \caption{Comparison of two switching-current criteria in the five-channel hybrid model. The global maximum of \(I_{\mathrm{tot}}\) does not necessarily coincide with the current at which one ML channel reaches its critical phase. The global maximum appears to provide a larger critical current and it happens when the all MLs are subcritical. The plateau originates from the ML switching.}
    \label{fig:Ic_defs}
\end{figure}

The distinction between these two criteria is illustrated in Fig.~\ref{fig:Ic_defs}. This figure is obtained from the CPR of the device shown in Fig.~\ref{fig:Ic_defs_cpr}.
The mathematical maximum identifies the largest value of the total current over the allowed phase window, whereas the channel-switching criterion identifies the current at which a particular metallic-link channel reaches its critical phase. 

Because the experimental switching event is likely triggered by one weak link becoming unstable, the ML critical phase criterion is more appropriate for interpreting the observed plateau-like features (Fig.~\ref{fig:deviced_I_sym_breaking}).

\subsection{Mathematical Origin of the Flat \(I_c(b)\) Plateaus}
\label{subsec:hybrid_flat_plateau}

The horizontal, or nearly horizontal, regions observed in the experimental switching envelope can be understood from the channel-switching criterion. 
Such field-independent switching occurs when the switching event is triggered by the middle ML channel, \(k=2\). 
Since this channel is chosen as the reference channel, its phase and critical phase are
\begin{equation}
    \phi_2=\phi_{\mathrm{ref}}.
    \label{eq:middle_wire_reference_phase}
\end{equation}
Let \(\Phi_s\) be the switching phase of the middle ML channel, which would be an integer multiple of \(\pi\), \(\Phi_s=\ell\phi_{c,2}\).

Using the phase relations above, the two SIS phases are
\begin{equation}
    \phi_1=\Phi_s-\frac{\pi b}{2}+2\pi v_1,
    \qquad
    \phi_3=\Phi_s+\frac{\pi b}{2}-2\pi v_2.
\end{equation}
The integer \(2 \pi\) vorticity offsets do not affect the sinusoidal SIS current because 
\begin{equation}
    \sin(\phi+2\pi N) = \sin\phi,\qquad \rm{for\ integer\ } N.
\end{equation}
Therefore the total current (\(I_{\mathrm{JJ,pair}}\)) carried by the two SIS channels is
\begin{equation}
\begin{aligned}
I_{\mathrm{JJ,pair}}
&=
I_{J,1}\sin\!\left(\Phi_s-\frac{\pi b}{2}\right)
+
I_{J,3}\sin\!\left(\Phi_s+\frac{\pi b}{2}\right) \\
&=
(I_{J,1}+I_{J,3})
\sin\Phi_s
\cos\!\left(\frac{\pi b}{2}\right) \\
&\quad
+
(I_{J,3}-I_{J,1})
\cos\Phi_s
\sin\!\left(\frac{\pi b}{2}\right).
\end{aligned}
\label{eq:hybrid_jj_pair_current_general}
\end{equation}
This expression shows that placing SIS channels symmetrically about the middle ML channel alone do not guarantee a flat plateau for arbitrary \(\Phi_s\). However, if the two SIS channels are similar in terms of critical current, then
\begin{equation}
    I_{\mathrm{JJ,pair}}
    =
    2I_{J,1}\sin\Phi_s\cos\!\left(\frac{\pi b}{2}\right),
    \qquad \text{when } I_{J,1}=I_{J,3}.
\end{equation}
Thus, in the ideal limit \(\Phi_s=\ell\pi\), the SIS contribution vanishes:
\begin{equation}
    I_{\mathrm{JJ,pair}} = 0
    \qquad
    \text{for}
    \qquad
    I_{J,1}=I_{J,3}, \quad \Phi_s\simeq \ell\pi
\end{equation}

Now consider the three metallic-link channels. For middle-channel switching, their contribution is
\begin{equation}
\begin{aligned}
    I_{\mathrm{ML}}^{(2)}
    &=
    K_0\phi_0+K_2\phi_2+K_4\phi_4 \\
    &=
    K_0\left[\Phi_s-\pi b+2\pi(v_0+v_1)\right] \\
    &\quad
    +K_2\Phi_s
    +K_4\left[\Phi_s+\pi b-2\pi(v_2+v_3)\right] \\
    &=
    (K_0+K_2+K_4)\Phi_s
    +\pi b(K_4-K_0) \\
    &\quad
    +2\pi\left[
        K_0(v_0+v_1)-K_4(v_2+v_3)
    \right].
\end{aligned}
\label{eq:hybrid_ml_current_middle_switching}
\end{equation}
The linear \(b\)-dependent term cancels exactly only when the two outer ML channels have equal phase stiffness,
\begin{equation}
    K_0 = K_4.
\end{equation}
If \(K_0\approx K_4\), the term does not vanish exactly, but it gives only a small residual slope proportional to \(\pi(K_4-K_0)\).
In the symmetric limit \(K_0=K_4=K_{\mathrm{out}}\), the metallic-link contribution becomes
\begin{equation}
    I_{\mathrm{ML}}^{(2)}
    =
    (2K_{\mathrm{out}}+K_2)\Phi_s
    +
    2\pi K_{\mathrm{out}}
    \left[
        (v_0+v_1)-(v_2+v_3)
    \right],
    \label{eq:hybrid_ml_current_middle_switching_symmetric}
\end{equation}
which is independent of \(b\).

Combining the SIS and metallic-link contributions, the full switching current for middle-channel switching is
\begin{equation}
\begin{aligned}
I_{\mathrm{sw}}^{(2)}
&=
(K_0+K_2+K_4)\Phi_s
+\pi b(K_4-K_0) \\
&\quad
+2\pi\left[
K_0(v_0+v_1)-K_4(v_2+v_3)
\right] \\
&\quad
+(I_{J,1}+I_{J,3})
\sin\!\Phi_s
\cos\!\left(\frac{\pi b}{2}\right) \\
&\quad
+(I_{J,3}-I_{J,1})
\cos\!\Phi_s
\sin\!\left(\frac{\pi b}{2}\right).
\end{aligned}
\label{eq:hybrid_middle_switching_current_general}
\end{equation}
Therefore, an exactly flat plateau occurs in the ideal symmetric limit 
\begin{equation}
    K_0=K_4,\qquad I_{J,1}=I_{J,3},\qquad \Phi_s=\ell\pi.
\end{equation}
Under these assumptions,
\begin{equation}
    I_{\mathrm{sw}}^{(2)}
    =
    (2K_{\mathrm{out}}+K_2)\ell\pi
    +
    2\pi K_{\mathrm{out}}
    \left[
        (v_0+v_1)-(v_2+v_3)
    \right],
    \label{eq:hybrid_flat_plateau_current}
\end{equation}
which contains no explicit dependence on \(b\). The vertical level of the plateau still depends on the vorticity imbalance between the left and right side of the device. Therefore different vorticity states can generate multiple nearly horizontal plateaus at different critical-current values, such as those seen in Fig.\ref{fig:deviced_I_sym_breaking}.

With small deviations from the ideal symmetry conditions, we pick up a residual linear slope, and thus experimental plateaus are generally not perfectly flat. If \(K_0\neq K_4\), the plateau acquires a residual linear slope proportional to \(\pi(K_4-K_0)\). If \(I_{J,1}\neq I_{J,3}\), the SIS pair contributes a residual field-dependent term proportional to \((I_{J,3}-I_{J,1})(-1)^\ell\sin(\pi b/2)\) when \(\Phi_s\simeq \ell\pi\). If \(\Phi_s\) is not close to an integer multiple of \(\pi\), even matched SIS channels contribute a field-dependent term proportional to \(\cos(\pi b/2)\). Thus the exactly flat plateau should be viewed as a limiting case. In the real device, small deviations from this limit produce residual slopes and weak field-dependent terms, rather than destroying the plateau completely.

\section{Discussion}
\subsection{Preservation of IB symmetry}
%more phase difference in the nanowire & increasing nanobridge length-> more negative peaks
%non-linear -> depletion of condensed states

Our observation of preserved IB symmetry in multi-weak-link nanowire devices is naturally explained by the multiple-nanowire SQUID (MW-SQUID) framework, in which several parallel superconducting nanowires are phase-coherent through two macroscopic superconducting electrodes. In this picture, the phase bias across each wire is not independent but is constrained by (i) Meissner screening currents in the electrodes under an applied perpendicular magnetic field and (ii) the integer vorticity (fluxoids) trapped in the cells between neighboring wires. This correlation is captured by the Meissner phase equation (MPE), which relates phase differences across different wires through a field-dependent term set by wire position and an integer vorticity term. In addition, each nanowire is assumed to obey an antisymmetric (odd) current--phase relation (CPR), typically approximated as linear at low temperature, and switching is governed by a critical-phase criterion: once any wire reaches its critical phase magnitude, the device undergoes a phase-slip event and the entire device typically transitions to the resistive state, due to the released heat. Together, these assumptions yield multi-valued $I_c(B)$ characteristics associated with distinct vorticity configurations, and define vorticity stability regions (VSRs) in the $I$--$B$ plane whose boundaries form the experimentally observed switching-current branches. Note that, as the current is reduced and the device returns back to the superconducting state, the new vorticity can be different from the vorticity in the previous cycle.

Within this model, IB symmetry emerges as a direct consequence of the antisymmetry of the CPR combined with the structure of the MPE. Specifically, the governing equations are invariant under simultaneous inversion of current and magnetic field, when accompanied by vorticity inversion (vortex $\leftrightarrow$ antivortex), implying that each VSR has a corresponding partner under $(I,B)\rightarrow(-I,-B)$. When we construct the experimentally relevant envelope currents, $I_{c,+}(B)$ and $I_{c,-}(B)$, by taking the upper and lower switching-current boundaries over all accessible vorticity states, this state-resolved symmetry reduces to the IB symmetry of the envelopes,
\begin{equation}
I_{c,+}(B)= -I_{c,-}(-B).
\end{equation}
Importantly, the same framework also highlights a practical diagnostic: if vortices are trapped inside the electrodes (rather than only in the inter-wire cells), they can introduce additional phase offsets that need not transform under $(I,B)\rightarrow(-I,-B)$, thereby violating IB symmetry. Thus, the robust IB symmetry observed in our multi-nanowire devices is consistent with a vortex-free electrode state and with switching dominated by coherent phase constraints rather than extrinsic rectification mechanisms. 

As was mentioned above, when the SQUID cools down, it enters a superconducting state with a random vorticity. Thus, multi-valued critical current versus magnetic field curves (distributions) are frequently observed, as a result of multiple measurements. They also obey IB-symmetry because if $B$ and $I$ are inverted the vorticity distribution obtained after the current ramp-down is typically inverted also.

For four-nanowire devices, the MW-SQUID model further predicts a particularly rich but still symmetry-protected structure of $I_c(B)$. A symmetric 4-wire SQUID (three cells) supports multiple distinct VSR geometries beyond the familiar diamond, including glider-shaped, trapezoidal, tilted-triangular, and kite-like regions, which can be indexed by the two-element vorticity-difference array $\Delta v=[v_{2,3}-v_{1,2},\,v_{3,4}-v_{1,2}]$. These diverse VSR boundaries naturally generate a non-sinusoidal, multi-lobe modulation of the switching-current envelope, providing a qualitative explanation for the highly structured $I_c(B)$ patterns observed experimentally while maintaining the envelope IB symmetry described above. Moreover, the generalized Little--Parks periodicity for an equidistant $n$-wire array predicts a fundamental period set by the number of cells, $\Delta b=n-1$; for $n=4$ this gives $\Delta b=3$, with principal maxima corresponding to ``equal-phase-bias'' conditions where all four wires can simultaneously reach their critical phases. In the idealized symmetric case, the next principal maximum occurs at $b=3$, which can be interpreted as the applied Meissner phase shift between neighboring wires being compensated by introducing one fluxoid in each of the three cells. The model also anticipates secondary maxima between principal maxima, associated with specific vorticity distributions, further enriching the envelope shape without requiring any symmetry breaking. 

Finally, the same four-wire analysis clarifies how disorder influences the detailed shape of $I_c(B)$ without necessarily destroying IB symmetry. Variations in critical phase or critical current can shift VSR vertices and distort VSR shapes while preserving the underlying periodicity in $b$ for fixed wire positions, whereas position disorder (unequal cell sizes) can change the period and, in incommensurate cases, render the response aperiodic. In addition, the model identifies a threshold in critical phase below which VSR sequences become topologically disjoint (for 4-wire devices, around $\phi_c\lesssim 3\pi/4$), producing field intervals where no superconducting state is stable and implying the possibility of full modulation and field-driven normal--superconducting transitions at $T\to 0$. The absence of such ``forbidden'' field windows in our symmetry-preserved devices suggests that our effective critical phases lie above this disjointness threshold, while the remaining sample-to-sample differences in modulation depth and fine structure can be attributed to realistic disorder in wire parameters and/or spacing within a framework that still enforces IB symmetry at the envelope level.

\subsection{Vorticity-State Selection and Symmetry Breaking.}

\begin{figure} [htbp]
    \centering
    \includegraphics[width=\linewidth]{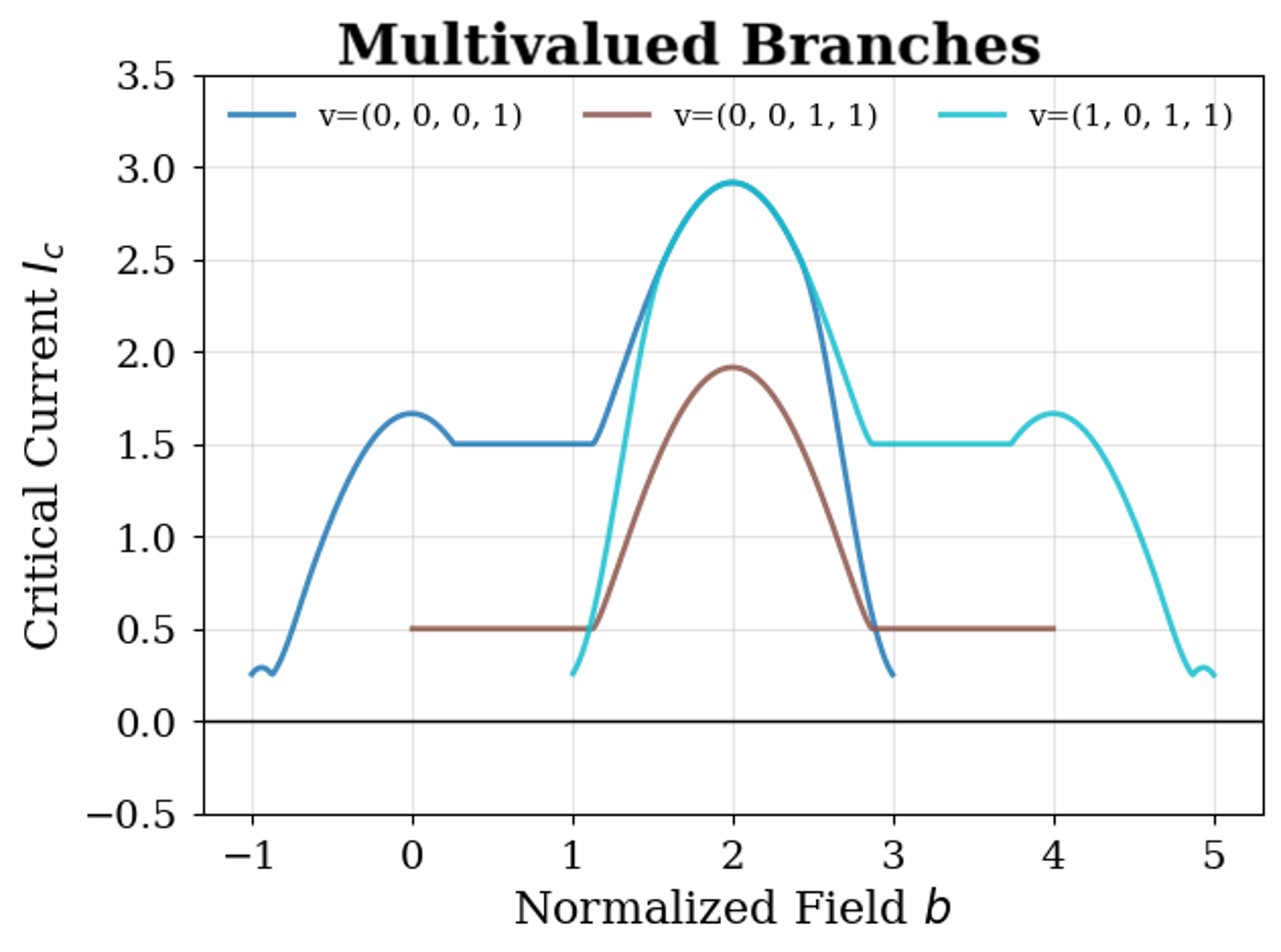}
    \caption{Simulated multivalued critical current branches for the positive currents. The global maximum criterion was used. These curves represent critical currents corresponding to low-energy vorticity states associated, namely $\mathbf v = (0, 0, 0, 1)$, $(1, 0, 1, 1)$, and $(0, 0, 1, 1)$.}
    \label{fig:multivalued_branches}
\end{figure}

The simulated multilevel switching-current structure shown in Fig.~\ref{fig:multivalued_branches} provides a useful comparison to the plateau-and-peak structure observed experimentally in Fig.~\ref{fig:deviced_I_sym_breaking}. 
In the simulation, the envelope is not produced by a single smooth interference branch, but by transitions among a small number of low-energy vorticity configurations. 
This supports the interpretation that the flat regions and peak-like features in Fig.~\ref{fig:deviced_I_sym_breaking} arise from changes in the selected vorticity state.

As shown in Fig.~\ref{fig:deviced_I_sym_breaking}, the switching-current envelope contains both nearly flat regions and pronounced peak-like features. 
We interpret the transitions between these two types of behavior as arising from changes in the vorticity configuration of the device.

To identify which vorticity states are energetically favored, we calculated the effective channel potential energy for each vorticity state:
\begin{equation}
    U_{\mathbf v}
    =
    \sum_{k\in\{0,2,4\}}
    \frac{\Phi_0 K_k}{4\pi}\phi_k^2
    +
    \sum_{k\in\{1,3\}}
    \frac{\Phi_0 I_{J,k}}{2\pi}
    \left[1-\cos\phi_k\right],
\end{equation}

\begin{figure} [htbp]
    \centering
    \includegraphics[width=\linewidth]{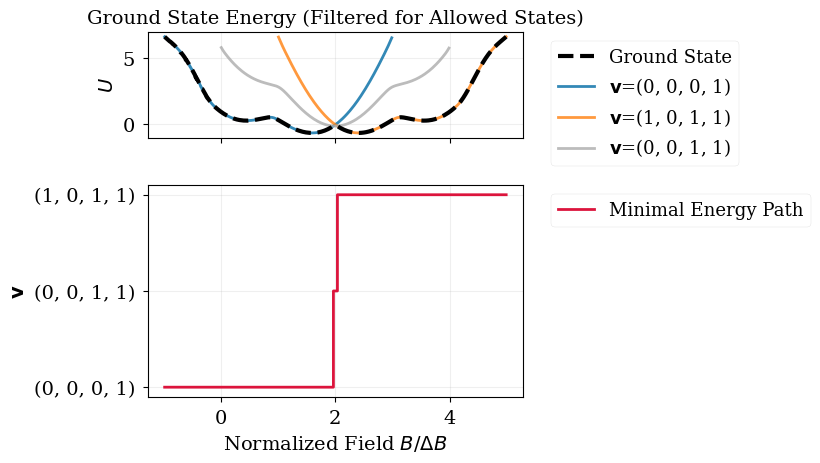}
    \caption{Ground-state energy and vorticity-state selection in the hybrid ML--SIS--ML--SIS--ML model. The calculated minimum-energy configuration evolves from $(0,0,0,1)$ to $(0,0,1,1)$ and then to $(1,0,1,1)$ as the normalized field \(b=B/\Delta B\) is varied.  The intermediate state $(0,0,1,1)$ is stable only near \(b\approx 2\), suggesting that the observed multilevel switching-current structure arises from transitions between discrete vorticity configurations \(\mathbf v\).}
    \label{fig:vorticity_energy}
\end{figure}

This effective energy includes the weak-link contributions and neglects additional loop-inductance and electrode kinetic-energy terms.
For each vorticity state, the ground-state energy is obtained by minimizing over the allowed reference-phase window,
\[
E_{\mathbf v}^{\mathrm{gs}}(b)
=
\min_{\phi_{\mathrm{ref}}\in\mathcal W(b,\mathbf v)}
U_{\mathbf v}(\phi_{\mathrm{ref}},b).
\]
The corresponding ground-state energy landscape is shown in Fig.~\ref{fig:vorticity_energy}. 
As the normalized field \(b=B/\Delta B\) is varied, the minimum-energy configuration changes between different vorticity states. 
This indicates that the experimentally observed multilevel envelope can arise from vorticity-state selection rather than from a single-valued SIS diffraction pattern.

Our numerical results indicate that the calculated response in Fig~\ref{fig:multivalued_branches} is composed of the three vorticity states—$\mathbf v = (0, 0, 0, 1)$, $(1, 0, 1, 1)$, and $(0, 0, 1, 1)$—that form competing low-energy configurations across the measured field range.

\begin{figure} [htbp]
    \centering
    \includegraphics[width=\linewidth]{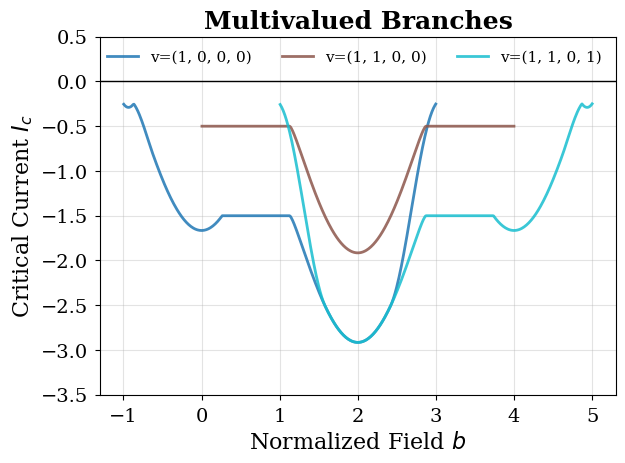}
    \caption{Simulated multivalued critical current branches for the negative current, $I_{c,-}$, over the same positive-field regime used in Fig.~\ref{fig:multivalued_branches}. The global minimum criterion was used. These curves show the negative-current counterparts of the low-energy vorticity states $\mathbf v = (1, 1, 0, 1)$, $(1, 0, 0, 0)$, and $(1, 1, 0, 0)$.}
    \label{fig:multivalued_branches_2}
\end{figure}

While the underlying hybrid equations retain the state-resolved \(IB\) symmetry when all partner vorticity states are accessible, the observed symmetry breaking suggests polarity-dependent vorticity selection. In this interpretation, the positive and negative current ramps access different subsets of metastable vorticity states, for example because vortex-entry or vortex-exit barriers are different for the two current polarities. We note that such situation is important in devices with larger critical currents, where the magnetic field of the bias current cannot be neglected and can impact the vorticity. The middle ML channel can produce a nearly flat switching level, while the outer ML channels and the vorticity state determine which level is selected.

The resulting ground-state energy landscape is shown in Fig.~\ref{fig:vorticity_energy} (black curve). 
As the normalized field \(b=B/\Delta B\) is swept, the minimum-energy configuration changes between different vorticity states. 
Specifically, the system evolves from the state $(0,0,0,1)$ to an intermediate state $(0,0,1,1)$, and then to $(1,0,1,1)$. 
The intermediate state $(0,0,1,1)$ is energetically favored only over a narrow field window near the center of the transition, approximately around \(b\approx 2\). 
This indicates that the experimentally observed multiple-segment structure can be interpreted as transitions between competing vorticity configurations, rather than as a single smooth critical-current function.
This picture provides a possible explanation for the symmetry-breaking behavior observed in Device D. 
The measured data suggest that the positive-current branch, $I_{c,+}$, preferentially follows the intermediate vorticity configuration over a finite field range, whereas the negative-current branch, $I_{c,-}$, more strongly favors one of the neighboring states, either $(0,0,0,1)$ or $(1,0,1,1)$. 
As a result, the two current polarities sample different portions of the multivalued energy landscape, leading to distinct switching-current envelopes after the transformation $I_{c,-}(B)\rightarrow -I_{c,-}(-B)$. 
In this interpretation, the observed $IB$-symmetry breaking does not simply arise from a rigid horizontal or vertical shift of a single envelope. 
Instead, it reflects polarity-dependent selection among metastable or low-energy vorticity states.

\section{Conclusion}

In this work, we investigated the symmetry properties of the magnetic-field-dependent critical current in superconducting nanodevices containing multiple superconducting channels with different types of CPR. We focused on the simultaneous inversion of bias current and magnetic field, referred to as $IB$ symmetry, expressed as
$I_{c,+}(B) = -I_{c,-}(-B)$. Across several nanowire-based devices, including Al and Ta nanowires fabricated on suspended SiN bridges and an Al-coated Ag nanowire, we observed that this symmetry is preserved even when the measured $I_c(B)$ patterns are highly non-sinusoidal, irregular, and multivalued. This result demonstrates that the preservation of $IB$ symmetry does not require a simple Fraunhofer-like interference pattern, but can remain robust in complex multi-weak-link superconducting networks.

The multiple-nanowire SQUID model predicts that IB-symmetry is valid for any random set of nanowire parameters. Our experiments confirmed that. In this model, the phase differences across the weak links are constrained by the applied magnetic field and by integer vorticity states between neighboring wires. When all accessible vorticity configurations are included, the maximum of the critical current remains symmetric under simultaneous inversion of current and magnetic field. Even if different vorticity configuration are realized in repetitive measurements, the resulting multi-valued function of the critical current of the magnetic field remains symmetric under simultaneous inversion of the bias current and the magnetic field. The observed symmetry preservation in Devices A--C supports these predictions. 

++++

In contrast, hybrid devices containing two types of Josephson junctions, namely SIS junctions (sinusoidal CPR) and metallic weak links (linear CPR), exhibit clear violations of $IB$ symmetry. In the Sn-based hybrid devices, we identify two forms of symmetry breaking: one in which the peak positions remain approximately fixed while the positive and negative switching-current magnitudes differ. In this case, I-symmetry is broken while B-symmetry is preserved. Another type is such that the two branches show both vertical mismatch and horizontal displacement in magnetic field, thus all types of symmetry (I and B and IB) are broken. 

To interpret this behavior, we developed a hybrid SIS--ML model containing both sinusoidal Josephson-junction current--phase relations and quasi-linear nanowire current--phase relations. The corresponding energy landscape combines periodic JJ terms with parabolic nanowire terms, allowing different vorticity states to become energetically favored over different magnetic-field ranges. This provides a possible microscopic mechanism for the observed symmetry breaking: the positive and negative current branches may follow different metastable or ground-state vorticity configurations during switching. In this picture, the broken $IB$ symmetry arises from polarity-dependent vorticity-state selection in a multivalued energy landscape.

Yet another type of the symmetry breaking  is observed near the central maximum in the Nb/BST superconducting island proximity array (Device E). In this case, the IB symmetry is broken only at weak magnetic fields, near zero, within the first .These results indicate that the topological nature of the conducting material, which is related to the spin-momentum correlation of the allowed electronic states, contributes to the IB symmetry breaking in these topological arrays.

Overall, our results show that $IB$ symmetry is a useful diagnostic for distinguishing coherent multi-vorticity superconducting interference from symmetry-broken hybrid weak-link behavior. The preservation of $IB$ symmetry in multi-nanowire devices highlights the robustness of phase-coherent vorticity physics, while its violation in hybrid SIS--ML and topological-insulator-based structures points toward mechanisms relevant to superconducting nonreciprocity and diode-like effects. These findings provide a framework for understanding when critical-current symmetry is protected, when it can be broken, and how engineered weak-link architectures may be used to control nonreciprocal superconducting transport in future superconducting and quantum devices.

%TC:ignore
\begin{acknowledgments}
The work was supported in part by the NSF DMR-2104757
and by the NSF OMA 2016136 Quantum Leap Institute for
Hybrid Quantum Architectures and Networks (HQAN). We also like to thank the University of Illinois Materials Research Laboratory for providing us with capabilities of sample fabrication and imaging.

\end{acknowledgments}

\section*{References}
\bibliography{references}
%TC:endignore
\end{document}